\documentclass[paper]{JHEP3} 

\usepackage{amsmath, amsbsy}
\usepackage{multirow, booktabs}
\usepackage{graphicx}

\newcommand{\Orderof}[1]{{\mathcal{O}\left(#1\right)}}
\newcommand{\GeV}{\ensuremath{\mathrm{GeV}}}
\newcommand{\TeV}{\ensuremath{\mathrm{TeV}}}
\newcommand{\seconds}{\ensuremath{\mathrm{s}}}
\newcommand{\MPl}{\ensuremath{\mathrm{M}_{\mathrm{P}}}}
\newcommand{\proton}{\ensuremath{\mathrm{p}}}

\newcommand{\bquark}{\ensuremath{\mathrm{b}}}
\newcommand{\antibquark}{\ensuremath{\bar{\mathrm{b}}}}
\newcommand{\electron}{\ensuremath{\mathrm{e}^-}}
\newcommand{\positron}{\ensuremath{\mathrm{e}^+}}

\newcommand{\gravitino}{\ensuremath{{\widetilde{G}}}}

\newcommand{\Bino}{\ensuremath{{\widetilde B}}}

\newcommand{\neutralino}{\ensuremath{{\widetilde \chi}^{0}_{1}}}

\newcommand{\mgr}{\ensuremath{m_{\widetilde{G}}}}

\newcommand{\mgravitino}{\ensuremath{\mgr}}

\newcommand{\NTP}{\ensuremath{\mathrm{NTP}}}
\newcommand{\TP}{\ensuremath{\mathrm{TP}}}

\newcommand{\equil}{\ensuremath{\mathrm{eq}}}

\newcommand{\freezeout}{\ensuremath{\mathrm{f}}}

\newcommand{\Reheating}{\ensuremath{\mathrm{R}}}
\newcommand{\TR}{\ensuremath{T_{\Reheating}}}

\newcommand{\taustau}{\ensuremath{\tau_{\widetilde{\tau}_1}}}

\newcommand{\monetwo}{\ensuremath{m_{1/2}}}
\newcommand{\mzero}{\ensuremath{m_{0}}}
\newcommand{\tanb}{\ensuremath{\tan{\beta}}}
\newcommand{\mgut}{\ensuremath{M_\mathrm{GUT}}}

\newcommand{\Xtau}{\ensuremath{X_{\tau}}}

\newcommand{\Atau}{\ensuremath{A_{\tau}}}

\newcommand{\mtau}{\ensuremath{m_{\tau}}}
\newcommand{\mLL}{\ensuremath{m_{\mathrm{LL}}}}
\newcommand{\mRR}{\ensuremath{m_{\mathrm{RR}}}}
\newcommand{\mstauone}{\ensuremath{m_{\stauone}}}
\newcommand{\mstautwo}{\ensuremath{m_{\stautwo}}}
\newcommand{\mstauL}{\ensuremath{m_{\stauL}}}
\newcommand{\mstauR}{\ensuremath{m_{\stauR}}}
\newcommand{\thetastau}{\ensuremath{\theta_{\stau}}}

\newcommand{\stauone}{\ensuremath{{\widetilde{\tau}_{1}}}}
\newcommand{\stautwo}{\ensuremath{{\widetilde{\tau}_{2}}}}
\newcommand{\stau}{\ensuremath{{\widetilde{\tau}}}}
\newcommand{\stauR}{\ensuremath{{\widetilde{\tau}_{\mathrm{R}}}}}
\newcommand{\stauL}{\ensuremath{{\widetilde{\tau}_{\mathrm{L}}}}}

\newcommand{\hhiggs}{\ensuremath{{h}^0}}
\newcommand{\Hhiggs}{\ensuremath{{H}^0}}
\newcommand{\Ahiggs}{\ensuremath{{A}^0}}
\newcommand{\Hphiggs}{\ensuremath{H}^+}
\newcommand{\Hmhiggs}{\ensuremath{H}^-}
\newcommand{\Hpmhiggs}{\ensuremath{H^{\pm}}}

\newcommand{\mtop}{\ensuremath{m_{t}}}

\newcommand{\mh}{\ensuremath{m_{\hhiggs}}}
\newcommand{\mH}{\ensuremath{m_{\Hhiggs}}}
\newcommand{\mA}{\ensuremath{m_{\Ahiggs}}}
\newcommand{\mHpm}{\ensuremath{m_{\mathrm{H}^{\pm}}}}

\newcommand{\Ystau}{\ensuremath{Y_{\widetilde{\tau}}}}
\newcommand{\Ystaueq}{\ensuremath{Y^{\mathrm{eq}}_{\widetilde{\tau}}}}
\newcommand{\YstaumO}{\ensuremath{Y^{\mathrm{m}\Omega}_{\widetilde{\tau}}}}

\newcommand{\mZ}{\ensuremath{M_{\mathrm{Z}}}}
\newcommand{\mW}{\ensuremath{M_{\mathrm{W}}}}

\newcommand{\ssqw}{\ensuremath{s^{2}_{W}}}

\newcommand{\sa}{\ensuremath{s_{\alpha}}}

\newcommand{\sapb}{\ensuremath{s_{\alpha+\beta}}}
\newcommand{\sbma}{\ensuremath{s_{\beta-\alpha}}}

\newcommand{\ca}{\ensuremath{c_{\alpha}}}

\newcommand{\cb}{\ensuremath{c_{\beta}}}
\newcommand{\ctwob}{\ensuremath{c_{2\beta}}}

\newcommand{\stwoth}{\ensuremath{s_{2\thetastau}}}

\newcommand{\csqth}{\ensuremath{c^{2}_{\thetastau}}}
\newcommand{\ctwoth}{\ensuremath{c_{2\thetastau}}}

\newcommand{\Lagrangian}{\ensuremath{\mathcal{L}}}
\renewcommand{\L}{\mathrm{L}}
\newcommand{\R}{\ensuremath{\mathrm{R}}}

\newcommand{\stauROT}{\ensuremath{R_{\stau}}}
\newcommand{\stauMAT}{\ensuremath{\mathcal{M}_{\stau}^2}}

\newcommand{\couptriLR}[3]{\ensuremath{\widetilde{C}[#1,#2,#3]}}
\newcommand{\couptri}[3]{\ensuremath{C[#1,#2,#3]}}
\newcommand{\couptriDL}[3]{\ensuremath{C^{\mathrm{DL}}[#1,#2,#3]}}

\newcommand{\sigmavof}[1]{\ensuremath{\langle \sigma_{{#1}}v \rangle}}
\newcommand{\sigmatot}{\ensuremath{\sigma_{\mathrm{tot}}}}
\newcommand{\sigmav}{\ensuremath{\langle {\boldsymbol\sigma} v \rangle}}
\newcommand{\pcm}{\ensuremath {P_{\mathrm{eff}}}}
\newcommand{\Tf}{\ensuremath {T_{\mathrm{f}}}}
\newcommand{\Tfone}{\ensuremath {T_{\mathrm{f1}}}}
\newcommand{\gstarS}{\ensuremath {g_{*S}}}

\title{Thermal relic abundances of long-lived staus}
\author{Josef Pradler\\
        Max-Planck-Institut f\"ur Physik,
        F\"ohringer Ring 6, D--80805 Munich, Germany\\
        E-mail: \email{jpradler@mppmu.mpg.de}}

\author{Frank Daniel Steffen\\
        Max-Planck-Institut f\"ur Physik,
        F\"ohringer Ring 6, D--80805 Munich, Germany\\
        E-mail: \email{steffen@mppmu.mpg.de}}
\preprint{arXiv:0808.2462\\MPP-2008-107}  
\date{\today}
\abstract{ 
  In supersymmetric models with a long-lived stau being the lightest
  Standard Model superpartner, 
  the stau abundance during primordial nucleosynthesis is tightly
  constrained.
  Considering the complete set of stau annihilation channels in the
  minimal supersymmetric Standard Model (MSSM) with real parameters
  for scenarios in which sparticle coannihilations are negligible,
  we calculate the decoupling of the lighter stau from the primordial
  plasma and identify processes which are capable to deplete the
  resulting stau abundance significantly.
  We find particularly efficient stau annihilation at the resonance of
  the heavy CP-even Higgs boson and for a lighter stau with a sizeable
  left--right mixing due to enhanced stau-Higgs couplings.
  Even within the constrained MSSM, we encounter both effects leading
  to exceptionally small values of the resulting stau abundance.
  Prospects for collider phenomenology are discussed
  and possible implications of our findings are addressed with
  emphasis on gravitino dark matter scenarios.
}

\begin{document} 

\section{Introduction}
\label{sec:introduction}

The appearance of the lighter stau $\stauone$ as the lightest Standard
Model superpartner---or lightest ordinary superpartner (LOSP)---is a
commonplace occurrence even in supersymmetric (SUSY) models with
restrictive assumptions on the SUSY breaking sector such as the
constrained minimal supersymmetric Standard Model (CMSSM).  If the
lightest supersymmetric particle (LSP) is assumed to be the LOSP, this
parameter region is not considered because of severe upper limits on
the abundance of massive stable charged particles~\cite{Yao:2006px}.
However, for example, in axino/gravitino LSP
scenarios~\cite{Pagels:1981ke,Borgani:1996ag,Bonometto:1989vh,Covi:1999ty,Steffen:2007sp}
and in scenarios with broken R
parity~\cite{Allanach:2003eb,Allanach:2006st,Allanach:2007vi,Takayama:2000uz,Buchmuller:2007ui},%
\footnote{In this work we assume that R-parity is conserved.}
 the $\stau$ LOSP becomes unstable and thereby a viable option.
Indeed, supersymmetric models with a long-lived $\stauone$ LOSP are
particularly promising for collider
phenomenology~\cite{Drees:1990yw,Nisati:1997gb,Feng:1997zr,Ambrosanio:2000ik,Buchmuller:2004rq,Hamaguchi:2004df,Feng:2004yi,Brandenburg:2005he,DeRoeck:2005bw,Martyn:2006as,Steffen:2006hw,Ellis:2006vu,Hamaguchi:2006vu}:
Since the $\stauone$ LOSP could escape the collider detector as a
quasi-stable muon-like particle, it can be associated with signatures
that are very different from the excess in missing energy expected in
neutralino LSP scenarios.

In the early Universe the negatively charged LOSP $\stauone$'s and the
associated positively charged anti-staus $\stauone^*$'s were in
thermal equilibrium for temperatures of $T>m_{\stauone}/20\gtrsim
T_{\freezeout}$. At $T_{\freezeout}$, the annihilation rate of the (by
then) non-relativistic $\stauone$'s becomes smaller than the Hubble
rate so that they decouple from the thermal plasma.  Thus, for
$T\lesssim T_{\freezeout}$, their yield
$Y_{\stau}\equiv(n_{\stauone}+n_{\stauone^*})/s$ is given
approximately by $\Ystau\approx Y^{\equil}_{\stau}(T_{\freezeout})$,
where $n_{\stau}^{(\equil)}\equiv n_{\stauone}^{(\equil)}+
n_{\stauone^*}^{(\equil)}$ is the (equilibrium) number density of both
$\stauone$ and $\stauone^*$ and $s=2\pi^2\,g_{*S}\,T^3/45$ the entropy
density with \gstarS\ effective degrees of freedom. This thermal relic
abundance $\Ystau$ is subject to cosmological constraints in SUSY
scenarios with a long-lived $\stauone$ LOSP:

\begin{itemize}

\item In axino/gravitino LSP scenarios, $\Ystau$ governs the
  non-thermally produced (NTP) relic density of axino/gravitino dark
  matter that originates from $\stauone$
  decays~\cite{Borgani:1996ag,Covi:1999ty}
        $\Omega_{\widetilde{\mathrm{a}}/\gravitino}^{\NTP} h^2
        = 
        m_{\widetilde{\mathrm{a}}/\gravitino}\, \Ystau\, s(T_0) h^2 / \rho_{\mathrm{c}}$
        where $m_{\widetilde{\mathrm{a}}/\gravitino}$ denotes the
        axino/gravitino LSP mass and $\rho_c/[s(T_0)h^2]=3.6\times
        10^{-9}\,\GeV$~\cite{Yao:2006px}. Thus, the dark matter
        density $\Omega_{\mathrm{dm}}$ which limits
        $\Omega_{\widetilde{\mathrm{a}}/\gravitino}^{\NTP}$ from above
        implies an upper limit on $\Ystau$ for a given
        $m_{\widetilde{\mathrm{a}}/\gravitino}$. This limit can become
        particularly restrictive in the case of additional sizeable
        contributions to $\Omega_{\mathrm{dm}}$ such as the ones from
        thermal axino/gravitino production
        $\Omega_{\widetilde{\mathrm{a}}/\gravitino}^{\TP}$~\cite{Bolz:2000fu,Brandenburg:2004du,Pradler:2006qh,Rychkov:2007uq}.
        For example, for
        $m_{\widetilde{\mathrm{a}}/\gravitino}=50~\GeV$ and
        $\Omega_{\widetilde{\mathrm{a}}/\gravitino}^{\TP}=0.99\,\Omega_{\mathrm{dm}}$
        ($0.9\,\Omega_{\mathrm{dm}}$), one finds $\Ystau<10^{-13}$
        ($10^{-12}$); cf.\ Fig.~13 of Ref.~\cite{Steffen:2006hw}.

      \item For $\stauone$ decays during/after big bang
        nucleosynthesis (BBN), the Standard Model particles emitted in
        addition to the axino/gravitino LSP can affect the abundances
        of the primordial light elements.  This leads to upper limits
        on $\xi_{\mathrm{em}/\mathrm{had}}\equiv
        \epsilon_{\mathrm{em}/\mathrm{had}}\,\Ystau$ that depend on
        the stau lifetime
        $\tau_{\stauone}$~\cite{Cyburt:2002uv,Kawasaki:2004qu,Jedamzik:2006xz}.
        Here $\epsilon_{\mathrm{em}/\mathrm{had}}$ denotes the
        (average) electromagnetic/hadronic energy emitted in a single
        $\stauone$ decay, which can be calculated with particle
        physics methods for a given model. Accordingly, the BBN
        constraints on $\xi_{\mathrm{em}/\mathrm{had}}$ can be
        translated into upper limits on $\Ystau$; cf.\ Fig.~12 of
        Ref.~\cite{Steffen:2006hw} (and Figs.~14 and~15 of
        Ref.~\cite{Kawasaki:2008qe}) for associated $\Ystau$ limits in
        gravitino LSP scenarios, which can be as restrictive as
        $\Ystau<10^{-14}$ ($10^{-15}$).

      \item The mere presence of the negatively charged $\stauone$'s
        at cosmic times of $t\gtrsim 5\times 10^3\,\seconds$ can lead
        to ($^4$He$\,\stauone$) and ($^8$Be$\,\stauone$) bound states
        and can thereby allow for catalyzed BBN (CBBN) of $^6$Li and
        $^9$Be to abundances far above the ones obtained in standard
        BBN
        (SBBN)~\cite{Pospelov:2006sc,Bird:2007ge,Pospelov:2007js,Pospelov:2008ta}.
        Indeed, confronting the abundances obtained in CBBN with
        observationally inferred bounds on the primordial abundances
        of $^9$Be (and $^6$Li) imposes restrictive upper limits of
        $\Ystau\lesssim 2\times 10^{-15}$ ($2\times
        10^{-15}$\,--\,$2\times 10^{-16})$ for $\tau_{\stauone}\gtrsim
        10^5\,\seconds$; cf.\ Fig.~5 in Ref.~\cite{Pospelov:2008ta}
        for $n_{\stauone}=n_{\stauone^*}$.

\end{itemize}

For example, in gravitino LSP scenarios with the $\stauone$ LOSP being
the next-to-lightest supersymmetric particle (NLSP) and conserved
R-parity, the listed cosmological constraints have been confronted
with representative
values~\cite{Asaka:2000zh,Fujii:2003nr}%
\begin{align}
  \label{eq:yield-approx}
  \Ystau
  \simeq 
  (0.4-1.5) \times 10^{-13} 
  \left( \frac{\mstauone}{100~\GeV} \right)
\ ,
\end{align}
which are in good agreement with the curves in Fig.~1 of
Ref.~\cite{Asaka:2000zh} that have been obtained for the case of a
purely `right-handed' $\stauone\simeq\stauR$ NLSP and a bino-like
lightest neutralino, $\neutralino\simeq\Bino$, with a mass of
$m_{\Bino}=1.1\,\mstauone$.  Thereby, it has been found that the
(C)BBN constraints impose the limit $\taustau\lesssim 5\times
10^3\,\seconds$~\cite{Pospelov:2006sc,Pospelov:2008ta} with severe
implications in the collider-friendly region of $\mstauone < 1~\TeV$:
(i)~The $\tau_{\stauone}$ limit disfavors the kinematical
determination of $\mgravitino$~\cite{Steffen:2006wx} and thereby both
the determination of the Planck scale at
colliders~\cite{Buchmuller:2004rq} and the method proposed to probe
the maximum reheating temperature $\TR$ at
colliders~\cite{Pradler:2006qh}.
(ii)~Within the CMSSM, the $\tau_{\stauone}$ limit implies an upper
limit on the reheating temperature of $\TR\lesssim
10^7~\GeV$~\cite{Pradler:2006hh,Pradler:2007is,Pradler:2007ar} that
disfavors the viability of thermal leptogenesis with hierarchical
heavy Majorana
neutrinos~\cite{Fukugita:1986hr,Davidson:2002qv,Buchmuller:2004nz,Blanchet:2006be,Antusch:2006gy}.
(iii)~The $\tau_{\stauone}$ limit can point to a CMSSM mass spectrum
which will be difficult to probe at the Large Hadron Collider
(LHC)~\cite{Cyburt:2006uv,Pradler:2006hh,Pradler:2007is,Pradler:2007ar}.
Indeed, the $\tau_{\stauone}$ limit
can be relaxed only with a significant reduction
of~(\ref{eq:yield-approx}) which has been presented explicitly so far
only for non-standard cosmological scenarios with a very low value
of~$\TR$~\cite{Takayama:2007du} or with late-time entropy production
after $\stauone$ decoupling and before
BBN~\cite{Pradler:2006hh,Hamaguchi:2007mp}.%
\footnote{Note that some implications of the $\tau_{\stauone}$ limit
  can be evaded not only by relaxing it but also by respecting it,
  e.g., R-parity violation can lead to $\tau_{\stauone}<5\times
  10^3\,\seconds$~\cite{Buchmuller:2007ui}.}

In this work we calculate the decoupling of the lighter stau from the
primordial plasma by taking into account the complete set of stau
annihilation channels in the MSSM with real parameters for SUSY
spectra for which sparticle coannihilation is negligible. Using our
own code for the computation of the resulting thermal relic stau
abundance $\Ystau$, we examine explicitly (i)~the effect of
left--right mixing of the lighter stau, (ii)~the effect of large
stau--Higgs couplings, and (iii)~stau annihilation at the resonance of
the heavy CP-even Higgs boson $H^0$.  We consider both the
``phenomenological MSSM'' (pMSSM) (see, e.g.,~\cite{Djouadi:2002ze})
in which the soft SUSY breaking parameters can be set at the weak
scale, and the CMSSM, in which the gaugino masses, the scalar masses,
and the trilinear scalar couplings are assumed to take on the
respective universal values $\monetwo$, $\mzero$, and $A_0$ at the
scale of grand unification $\mgut\simeq 2\times 10^{16}\,\GeV$. Within
the framework of the pMSSM, we show examples in which $\Ystau$ can be
well below $10^{-15}$.  Even within the CMSSM, we encounter regions
with exceptionally small values of $\Ystau\lesssim 2\times 10^{-15}$.
The implications of these findings are discussed for scenarios with
the gravitino LSP and the stau NLSP.  We also address the viability of
a $\stauone$--$\stauone^*$ asymmetry.  Remarkably, we find that key
quantities for the significant $\Ystau$ reduction could be probed at
both the LHC and the International Linear Collider (ILC).

A calculation of the thermal relic abundance of long-lived staus has
also been part of a recent thorough study~\cite{Berger:2008ti} which
focusses on gauge interactions and on the effect of Sommerfeld
enhancement. In contrast, the most striking findings of our study---in
which Sommerfeld enhancement is not taken into account---are related
to the Higgs sector of the MSSM.  At this point, we should also stress
that the \texttt{micrOMEGAs}
code~\cite{Belanger:2001fz,Belanger:2004yn,Belanger:2007zz,Belanger:2008sj}
allows for sophisticated calculations of the thermal relic stau
abundance also in regions in which coannihilation effects become
important.  In fact, \texttt{micrOMEGAs} has already been applied in
several studies to calculate
$\Ystau$~\cite{Fujii:2003nr,Pradler:2006hh,Kersten:2007ab,Pradler:2007ar,Berger:2008ti}.
In this paper, we also work with \texttt{micrOMEGAs} to cross check
the results of our own $\Ystau$ calculation and to calculate $\Ystau$
in parameter regions in which sparticle coannihilations become
relevant.

The outline of this paper is as follows.  In the next section we
review basic properties of the staus to introduce our notations and
conventions for the stau mixing angle.  Section~\ref{sec:prim-ann}
explains the way in which we calculate $\Ystau$ and provides the
complete list of stau annihilation channels.  In
Sect.~\ref{sec:dependence} we analyze the dependence of the most
relevant stau annihilation channels on the stau mixing angle.  Effects
of large stau--Higgs couplings and stau annihilation at the $H^0$
resonance are studied in Sects.~\ref{sec:enhanc-coupl-higgs}
and~\ref{sec:reson-annih}, respectively.  The viability of a
$\stauone$--$\stauone^*$ asymmetry is addressed in
Sect.~\ref{sec:comm-stau-stau}.  In Sect.~\ref{sec:annih-chann} we
present exemplary parameter scans within the CMSSM that exhibit
exceptionally small $\Ystau$ values.  Potential collider phenomenology
of the parameter regions associated with those exceptional relic
abundances and potential implications for gravitino dark matter
scenarios are discussed in Sects.~\ref{sec:collider}
and~\ref{sec:gravitino}, respectively.

\section{Stau mixing and mass eigenstates}
\label{sec:properties-stau}

In this section we review some basic properties of the stau to set the
notation. In absence of inter-generational mixing, the stau
mass-squared matrix in the basis of the gauge eigenstates $(\stauL,
\stauR)$ reads
\begin{equation}
    \label{eq:stau-mass-matrix}
    \stauMAT = 
    \begin{pmatrix}
    \mtau^2 + \mLL^2 & \mtau \Xtau^* \\
             \mtau \Xtau      & \mtau^2 + \mRR^2      
    \end{pmatrix} = 
    (\stauROT)^\dagger
    \begin{pmatrix}
    \mstauone^2 & 0\\
    0 & \mstautwo^2   
    \end{pmatrix}
    \stauROT
\end{equation}
with
\begin{eqnarray}
    \label{eq:M-mstau-entries}
    \mLL^2 & = & \mstauL^2 +  \left( - \frac{1}{2} + \sin^2{\theta_W}  \right) \mZ^2 \cos{2\beta}\\
    \mRR^2 & = & \mstauR^2 - \sin^2{\theta_W}  \mZ^2 \cos{2\beta} \\
    \Xtau & = &\Atau - \mu^* \tanb \ .
\end{eqnarray}
Here, \mstauL\ and \mstauR\ are the soft SUSY breaking masses, \Atau\
is the trilinear coupling, $\mu$ is the Higgs-higgsino mass parameter,
and $\tanb=v_2/v_1$ denotes the ratio of the two Higgs vacuum expectation
values. In this work we restrict ourselves to the MSSM with real
parameters. Then $\Xtau^* = \Xtau$ so that the mass eigenstates
\stauone\ and \stautwo\ are related to \stauL\ and \stauR\ by means of
an orthogonal transformation
\begin{equation}
    \label{eq:phys-stau-fields}
    \begin{pmatrix}
      \stauone \\ \stautwo
    \end{pmatrix}
    = \stauROT
    \begin{pmatrix}
      \stauL \\
      \stauR
    \end{pmatrix}
    \quad\textrm{with}\quad
      \stauROT =
    \begin{pmatrix}
      \cos{\thetastau} & \sin{\thetastau} \\
      -\sin{\thetastau} & \cos{\thetastau}
    \end{pmatrix}
\end{equation} 
with \thetastau\ denoting the stau mixing angle. Imposing the mass
ordering $\mstauone < \mstautwo$ and choosing $0\leq \thetastau < \pi
$, the mixing angle can be inferred from the elements of \stauMAT,
\begin{equation}
    \label{eq:thetastau}
    \tan{2 \thetastau}  =  \frac{ 2 \mtau \Xtau }
    {  \mLL^2 - \mRR^2 } = \frac{ 2 \mtau \Xtau }{\delta} \ , \qquad
    \sin{2\thetastau} = \frac{ 2 \mtau \Xtau}{\mstauone^2 - \mstautwo^2} \ ,
\end{equation}
where the sign of the second relation determines the quadrant of
$\thetastau $. In the first relation, we have introduced $\delta
\equiv \mLL^2 - \mRR^2$. In particular, $\thetastau=\pi/2$ corresponds
to a purely right-handed stau, $\stauone=\stauR$,
whereas 
maximal mixing occurs for $\thetastau=\pi/4$ and $3\pi/4$.  The
physical stau masses are then given by
\begin{equation}
   \label{eq:stau-masses}
   m_{\stau_{1,2}}^2 = \mtau^2 + \mRR^2 +  \frac{1}{2} \left[\delta \mp
      \sqrt{\delta^2 + 4\mtau^2\Xtau^2 } \right]
\end{equation}
from which we see that an increase of $|\Xtau|$ leads to a reduction
of \mstauone.

\section{Calculation of the thermal relic stau abundance}
\label{sec:prim-ann}

We have undertaken the effort to set up our own full-fledged relic
abundance calculation. Let us in the following give a description of
our approach to compute the stau yield $\Ystau$.  Throughout this work
we assume a standard cosmological history with a temperature $T$ of
the primordial plasma above the stau decoupling temperature $\Tf$ so
that the lighter stau $\stauone$ was once in thermal
equilibrium. Then, the total stau yield $\Ystau\equiv Y_{\stauone} +
Y_{\stauone^*} $ is found by solving the well-known Boltzmann equation
\begin{align}
  \label{eq:boltzmann-equation}
   \frac{d\Ystau}{dt} = - s \sigmav
   \left[ \Ystau^2 - {(\Ystaueq)}^2 \right]\ .
\end{align}
Using the Maxwell--Boltzmann approximation, the stau equilibrium yield
\Ystaueq\ is given by
\begin{align}
   \label{eq:stau-eq}
   \Ystaueq = \frac{ \mstauone^2 T}{\pi^2 s}\, K_{2}\left(\frac{\mstauone}{T}\right)
\end{align}
and the thermally averaged annihilation cross section by~\cite{gondolo:1990dk}
\begin{align}
  \label{eq:thavg}
  \sigmav(T) =
  \frac{1}{2\mstauone^4 T [K_2(\mstauone/T)]^2}\,
  \int^\infty_{4\mstauone^2} ds \, 
  \sqrt{s} K_1\left(\frac{\sqrt{s}}{T}\right) \pcm^2 {\boldsymbol \sigma}(s) \,
  ,
\end{align}
where $K_i$ is the modified Bessel function of order $i$ and
$\pcm=\sqrt{s-4\mstauone^2}\Big/2$.

Note that \sigmav\ contains all the information from
the particle physics side. It is obtained by computing the total
stau-annihilation cross section,
\begin{align}
 \label{eq:sigmatot}
        {\boldsymbol \sigma} \equiv \frac{1}{2}\, \sigmatot
        \quad \mathrm{with} \quad
        \sigmatot =  \sigma_{\stauone\, \stauone
        \rightarrow \tau\tau } + \sum_X \sigma_{\stauone\, \stauone^*
        \rightarrow X }    
\ ,
\end{align}
where the sum for the annihilation of $\stauone\, \stauone^*$ pairs%
\footnote{Counting wise we distinguish between $\stauone\, \stauone^*
  \rightarrow X$ and the conjugate process $\stauone^*\, \stauone
  \rightarrow \overline{X}$. In absence of CP violation in the SUSY
  sector, their cross sections agree so that we can solve a single
  Boltzmann equation~(\ref{eq:boltzmann-equation}) for obtaining
  $\Ystau$.}
has to be taken over all final states $X$. The factor $1/2$ is
convention but gives (\ref{eq:boltzmann-equation}) its familiar form.
The complete list of annihilation processes in the MSSM with real
parameters---save for coannihilation processes---is given in
Table~\ref{tab:ann-channels}.%
\footnote{For a purely right-handed stau $\stauone=\stauR$, the stau
  annihilation channels and associated cross sections have already
  been presented in Ref.~\cite{Ellis:1999mm} in the context of
  $\neutralino$-$\stauone$ coannihilation.}
 In addition, this table shows all
possible particle exchanges, where $s$, $t$, and $u$ are the
Mandelstam variables which denote the respective channel. A number of
annihilation processes proceeds also via a four-point vertex. Those
are marked in the column named ``contact.'' Already by mere optical
inspection, we immediately see that the Higgs sector plays potentially
an important role in the determination of the stau yield $\Ystau$.

\begin{table}[t]
  \caption[Ann-ch]{The complete set of stau annihilation channels in the MSSM with real parameters for scenarios in which sparticle coannihilations are negligible. The mass eigenstates of the Higgs fields are denoted by $\hhiggs$, $\Hhiggs$, $\Ahiggs$, and $\Hpmhiggs$ and the ones of the neutralinos, the charginos, and the tau sneutrino by $\widetilde{\chi}^0_{1,..,4}$, $\widetilde{\chi}^{\pm}_{1,2}$, and $\widetilde{\nu}_{\tau}$, respectively. Because of the absence of a $\stauone\stauone\Ahiggs$ coupling (cf.\ Sect.~\ref{sec:enhanc-coupl-higgs}), $s$-channel exchange of the CP-odd Higgs boson $\Ahiggs$ and also $\stauone\stauone^*\rightarrow\gamma\Ahiggs$ do not appear.} 
\label{tab:ann-channels}
\begin{center}
\begin{tabular}{llc@{\qquad}c@{\qquad}c}
\toprule 
$\stauone^{(*)}\,\stauone^{(*)} \rightarrow$& final state  &  $s$-channel &
$t(u)$-channel & contact\\
\midrule 
\vspace{-0.4cm} \\ 
   & $\tau\tau 
    \ (\overline{\tau}\overline{\tau})
   $ & --- & $\widetilde{\chi}^0_{1,..,4}$ &---\\
   \vspace{-0.4cm} \\
   \midrule
   $\stauone\stauone^* \rightarrow$& final state $X$${}^\dagger$ &  $s$-channel
   &
   $t(u)$-channel & contact\\ 
   \midrule
   \vspace{-0.4cm} \\ 
   & $\mu\overline{\mu}$, $e\overline{e}$ &
   $\hhiggs,\Hhiggs,\,\gamma,Z$ 
   & --- & ---\\\vspace{-0.4cm} \\ 
   & $\tau\overline{\tau}$ &
   $\hhiggs,\Hhiggs,\,\gamma,Z$ 
   & $\widetilde{\chi}^0_{1,..,4}$ & ---\\\vspace{-0.4cm} \\ 
   & $\nu_{e}\overline{\nu}_{e}$, $\nu_{\mu}\overline{\nu}_{\mu}$  & $Z$ & --- & ---\\\vspace{-0.4cm}\\
   & $\nu_{\tau}\overline{\nu}_{\tau}$  & $Z$ & $\widetilde{\chi}^{\pm}_{1,2}$ & ---\\\vspace{-0.4cm}\\
   & $q_k\overline{q}_k$ & $\hhiggs,\,\Hhiggs,\,\gamma,\, Z$ & ---
   & --- \\\vspace{-0.4cm}\\
   & $\gamma\gamma,\, \gamma Z$ & --- & $\stauone$ & $\checkmark$ \\\vspace{-0.4cm}\\
   & $ZZ$ & $\hhiggs,\Hhiggs$ & $\stau_{1,2}$ & $\checkmark$ \\\vspace{-0.4cm}\\
   & $W^+W^-$ & $\hhiggs,\,\Hhiggs,\,\gamma,\,Z$ &
   $\widetilde{\nu}_{\tau}$ & $\checkmark$ \\\vspace{-0.4cm}\\
   & $\gamma\hhiggs,\,\gamma\Hhiggs$ & --- & $\stauone$ & ---\\\vspace{-0.4cm}\\
   & $Z\hhiggs,\,Z\Hhiggs$ & $Z$  &  $\stau_{1,2}$ & ---\\\vspace{-0.4cm}\\
   & $Z\Ahiggs$ & $\hhiggs,\Hhiggs$  &  $\stau_{2}$ & ---\\\vspace{-0.4cm}\\
   & $W^{\mp}\Hpmhiggs$ & $\hhiggs,\Hhiggs$  &  $\widetilde{\nu}_{\tau}$ & ---\\\vspace{-0.4cm}\\
   & \parbox{2.5cm}{
     $\hhiggs\hhiggs , 
     \, \hhiggs\Hhiggs,$\\
     $\Hhiggs\Hhiggs$}
   & $\hhiggs, \Hhiggs$& $\stau_{1,2} $ & $\checkmark$
   \\\vspace{-0.4cm}\\\vspace{-0.4cm}\\
   & \parbox{2.5cm}{
     $\Ahiggs\Ahiggs$}
   & $\hhiggs, \Hhiggs$& $\stau_{2} $ & $\checkmark$
   \\\vspace{-0.4cm}\\\vspace{-0.4cm}\\
   &  $\hhiggs\Ahiggs,\,\Hhiggs\Ahiggs$ & $Z$ &
   $\stau_{2} $  & ---\\\vspace{-0.4cm}\\
   & $\Hphiggs\Hmhiggs$ & $\hhiggs,\,\Hhiggs,\,\gamma,\,Z$
   & $\widetilde{\nu}_{\tau}$& $\checkmark$\\\vspace{-0.4cm}\\
   \bottomrule
   \multicolumn{4}{l}{\footnotesize ${}^\dagger$  $k=u,\,d,\,c,\,s,\,t,\,b$}\\[-0.1cm]
\end{tabular}
\end{center}
\end{table}

For all channels in Table~\ref{tab:ann-channels}, we generate
\texttt{Fortran} code for the squared matrix elements
$|\mathcal{M}_i|^2$ by using the computer algebra packages
\texttt{FeynArts~5.4}~\cite{Hahn:2000kx,Hahn:2001rv} and
\texttt{FormCalc~5.3}~\cite{Hahn:1998yk,Hahn:2006qw}. For a chosen
point in the SUSY parameter space, we then compute the radiatively
corrected superparticle spectrum by running the spectrum generator
\texttt{SuSpect 2.40}~\cite{Djouadi:2002ze}. Its output allows us to
set all SUSY parameters so that we can compute the total cross section
$\sigmatot(s)$ given by (\ref{eq:sigmatot}) and subsequently the
thermally averaged cross section~(\ref{eq:thavg}).
Numerically, the computation of (\ref{eq:thavg}) is the most demanding
part in the relic abundance calculation. In particular, we take special
care about the following cases:
\begin{itemize}
\item \Hhiggs -resonance: Resonant stau annihilation via $\Hhiggs$
  exchange is one of the central points in this paper. In the
  generation of the matrix elements, we have therefore included the
  total \Hhiggs-width $\Gamma_{\Hhiggs}$ in the respective $s$-channel
  propagators.

\item Propagator poles: A diverging $t(u)$-channel propagator can be
  encountered when a production threshold is met. We overcome this
  problem by including a ``sparticle-width'' of $0.01\mstauone$ in the
  respective propagators in the vicinity of dangerous thresholds. A
  particularly interesting example with a diverging $t(u)$-channel
  propagator is given by the process $\stauone\,\stauone^*\rightarrow
  \gamma\Hhiggs$ if $\sqrt{s}=m_{\Hhiggs}$ is fulfilled since then the
  \Hhiggs-exchange in the $s$-channels of other processes is resonant
  simultaneously.

\item Bessel functions: The Bessel functions in (\ref{eq:stau-eq}) and
  (\ref{eq:thavg}) exhibit an exponential behavior for large
  arguments $x\gg 1$~\cite{abramowitz+stegun}
  \begin{align}
    \label{eq:bessel}
    K_n(x) \simeq \sqrt{\frac{\pi}{2\,x}} e^{-x} 
    \left( 
      1 + \frac{4n^2-1}{8x} + \dots 
    \right)\ .
  \end{align}
  For small temperatures $T$, the arguments of $K_1$ and $K_2$ in
  (\ref{eq:thavg}) become large simultaneously. Therefore, in order to
  ensure numerical stability, we expand the Bessel functions in
  (\ref{eq:thavg}) for $\mstauone/T>35$ as in (\ref{eq:bessel}) and
  cancel the exponents analytically.%
  \footnote{J.P.\ is grateful to P.~Gondolo and J.~Edsjo for pointing
    out that trick.}
 \end{itemize}

 We find the starting point for the numerical integration of
 (\ref{eq:boltzmann-equation}) by solving~\cite{Belanger:2004yn}
 \begin{align}
   \label{eq:Tf1}
   \left. \frac{d\Ystaueq}{dT}\right|_{\Tfone} = \sqrt{\frac{8\pi^2
       g_{*}(T) }{45}} \MPl 
   \sigmav(\Ystaueq  )^2 \lambda(\lambda+2)
 \end{align} 
 where $g_{*}(T)$ is a degrees of freedom
 parameter~\cite{gondolo:1990dk} and $\MPl=2.4\times 10^{18}\ \GeV$
 the (reduced) Planck mass.  $\Tfone$ marks the point at which the
 stau starts to decouple chemically from the background plasma,
 $\Ystau(\Tfone)-\Ystaueq(\Tfone)\simeq \lambda \Ystaueq(\Tfone)$ with
 $\lambda = 0.1$~\cite{Belanger:2004yn} chosen in our code.  Since we
 use a globally adaptive Gaussian integration routine to calculate
 (\ref{eq:thavg}), the computation of $\sigmav(T)$ is time-demanding.
 Therefore, we evaluate (\ref{eq:thavg}) on a grid of different
 temperatures and use cubic spline interpolation to obtain values in
 between.
 We then solve the Boltzmann equation (\ref{eq:boltzmann-equation}) by
 numerical integration from $\Tfone$ to zero.  There, we fully take
 into the account the temperature dependence of $g_{*}$ and \gstarS\
 by interpolating the respective tabulated values provided as part of
 the relic density code \texttt{DarkSUSY~4.00}~\cite{gondolo:2004sc}.
 The freeze out temperature can then be defined by $\Tf\equiv
 (\Tfone+T_{\mathrm{f}2})/2$ where $T_{\mathrm{f}2}$ is given by
 $\Ystaueq(T_{\mathrm{f}2}) =
 \Ystau(T_{\mathrm{f}2})/10$~\cite{Belanger:2004yn}. For
 $T<T_{\mathrm{f}2}$, residual annihilations will further reduce
 $\Ystau$ so that we refer to the decoupling yield
 $\Ystau^{\mathrm{dec}}$ as the quantity at the endpoint of
 integration. For simplicity, we call this yield $\Ystau$ in the
 Introduction and in the following. Moreover, we will quantify $T$ in
 terms of $x\equiv m_{\stauone}/T$ and in particular $\Tf$ in terms of
 $x_{\freezeout}\equiv m_{\stauone}/\Tf$.

 Note that we have additionally modified the \texttt{FeynArts} MSSM
 model file for the generation of the matrix elements in two ways: The
 first version, which we use throughout
 Sects.~\ref{sec:dependence}--\ref{sec:reson-annih}, allows us to set
 all $q_k\overline{q}_k$--Higgs and all trilinear Higgs couplings by
 using the computer tool
 \texttt{FeynHiggs~2.6.3}~\cite{Heinemeyer:1998yj}; see also
 Sects.~\ref{sec:enhanc-coupl-higgs} and~\ref{sec:reson-annih}. The
 second version allows for a direct comparison with the existing
 computer code \texttt{micrOMEGAs 2.0.6}
 \cite{Belanger:2001fz,Belanger:2004yn,Belanger:2006is}. We have
 transcribed their routine~\cite{Pukhov:2004ca} for the computation of
 the running quark masses to \texttt{Fortran}, adopted all
 $q_k\overline{q}_k$--Higgs couplings, and modified all Higgs-self
 couplings of our matrix elements to match with their implemented
 version of the MSSM~\cite{Dubinin:1998nt}. Using this second version,
 we find perfect agreement between our codes.%
 \footnote{For our computation we use the Standard Model parameters
   $m_{\mathrm{t}}=172.5~\GeV$,
   $m_{\mathrm{b}}(m_{\mathrm{b}})^{\mathrm{\overline{MS}}} = 4.25\
   \GeV$, $\alpha_{\mathrm{s}}^{\mathrm{\overline{MS}}}(\mZ)=0.1172$,
   $\alpha_{\mathrm{em}}^{-1\mathrm{\overline{MS}}}(\mZ) = 127.932 $,
   and $\mZ=91.187\ \GeV$. Since \texttt{micrOMEGAs} has hard-coded
   $\sin{\theta_W}=0.481$ from which it computes $\mW$ using the
   on-shell relation with $\mZ$, we follow their convention to allow
   for a better comparison of our results with \texttt{micrOMEGAs}.}

 \section{Dependence of stau annihilation on the stau mixing angle}
\label{sec:dependence}

In order to isolate the distinct features of the different
annihilation processes we need to have full control over the
superparticle mass spectrum. Therefore, in the following, we will not
rely on any constrained model (such as the CMSSM) where the soft-SUSY
breaking parameters are subject to stringent boundary conditions at
some high scale (such as \mgut). In those models, the mass spectrum is
found only after renormalization group (RG) evolution from the high
scale down to the electroweak scale.  Instead, we choose to work in
the framework of the ``phenomenological MSSM'' (pMSSM), see,
e.g.,~\cite{Djouadi:2002ze}.  There, all soft-SUSY breaking parameters
can be set at the scale of electroweak symmetry breaking---a low
scale---which we fix to $\sim 2 \mstauone$. In particular, one can
also trade the Higgs mass-squared parameters $m_{H_u}^2$ and
$m_{H_d}^2$ against $\mu$ and the pseudoscalar Higgs boson mass
$m_{\Ahiggs}$.%
\footnote{Though the advocated procedure may require fine-tuning in
  the electroweak symmetry breaking conditions, it conveniently
  provides us with running parameters at the scale of stau
  annihilation.}
Choosing $\mu$ as an input parameter is very convenient for two
reasons: First, together with the specification of the gaugino masses
$M_{1,2}$ we have control over the gaugino/higgsino mixture of the
neutralinos $\widetilde{\chi}_i^0$. Second, $\mu$ enters directly into
the stau-Higgs couplings, whose importance will become clear in the
next section. Furthermore, in the following, we choose to set all
soft-SUSY breaking scalar masses (apart from \mstauL and \mstauR) to a
common value $M_S=1\ \TeV$. Thereby, we essentially decouple all
sfermions which are not of interest for us. This ensures also that we
never enter \emph{accidentally} any coannihilation regime.  Finally,
for simplicity, we set also all trilinear parameters to a common value
$A$. Given $\mu$, $\Atau=A$, and $\tanb$, and thereby $\Xtau$, we can
then fix $\mstauone$ and $\thetastau$ to arbitrary values by adjusting
$\mRR^2$ and $\delta$ in Eqs.~(\ref{eq:thetastau})
and~(\ref{eq:stau-masses}).

In the following, we will focus on two distinct regions of the SUSY
parameter space. In the beginning, we will choose $m_{\Ahiggs}$ to be
very large $m_{\Ahiggs}=1\ \TeV \gg \mZ$. This corresponds to the
decoupling limit of the MSSM where the following (tree-level)
relations hold~\cite{Gunion:2002zf}
\begin{align}
    \label{eq:DL-higgs-masses}
    \mh^2 &\simeq \mZ^2 \cos^{2}{2\beta}, \qquad \mH^2 \simeq \mA^2 +
    \mZ^2 \sin^{2}{2\beta},\\
    \label{eq:DL-cbma}   
    \mHpm^2 & = \mA^2 + \mW^2, \qquad \cos^{2}{(\beta-\alpha)} \simeq
    \frac{\mZ^4\sin^2{4\beta}}{4 \mA^4}.
\end{align}
Therefore, $\mA \simeq \mH \simeq \mHpm$ up to corrections
$\Orderof{\mZ^2/\mA}$ so that any of the stau annihilation channels
into heavy Higgs bosons is kinematically blocked. Furthermore,
$\cos{(\beta-\alpha)}=0$ up to corrections $\Orderof{\mZ^2/\mA^2}$
implies that the $\Hhiggs VV$ coupling ($V=Z,W$) becomes very small so
that we loose the $\Hhiggs$-exchanges in the stau annihilation
channels with a $VV$ final state. At the same time, the light Higgs
boson takes on its Standard Model value for the $\hhiggs VV$ coupling.
Complementary to that we will consider also regions of the SUSY
parameter space with smaller $\mA$, e.g., in the next section, where
we will put a stronger focus on the Higgs sector and its connection to
\Ystau.
%

\begin{figure}[t]
\begin{center}
\centerline{\includegraphics[height=7.2cm]{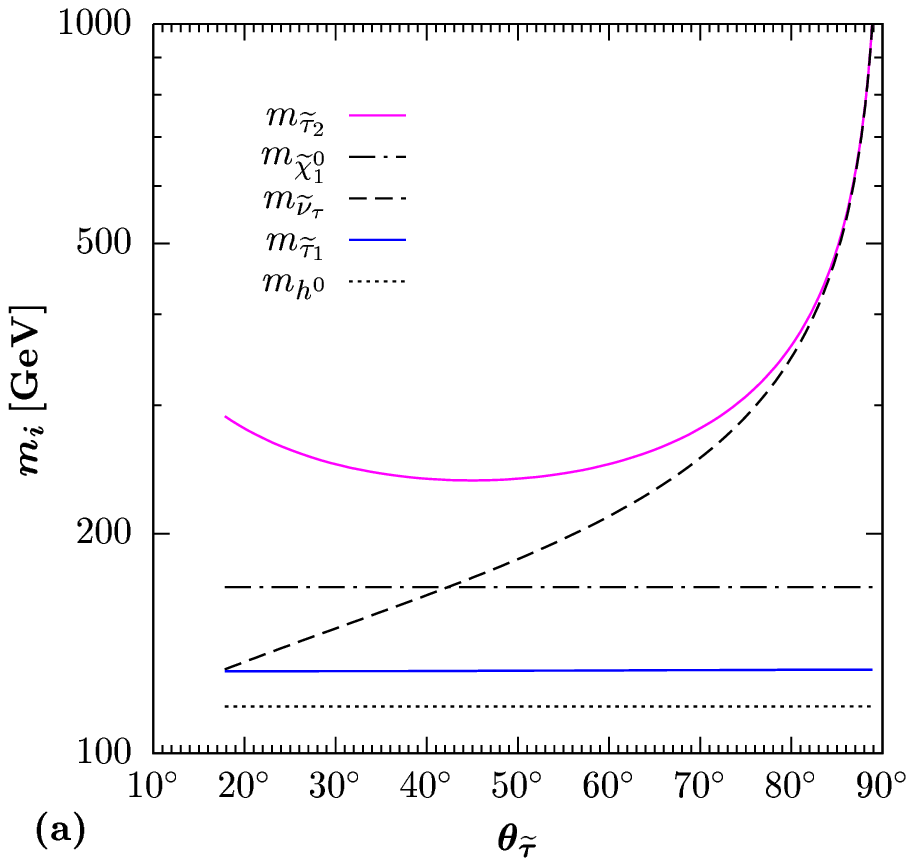}%
\hskip 0.0cm
\includegraphics[height=7.2cm]{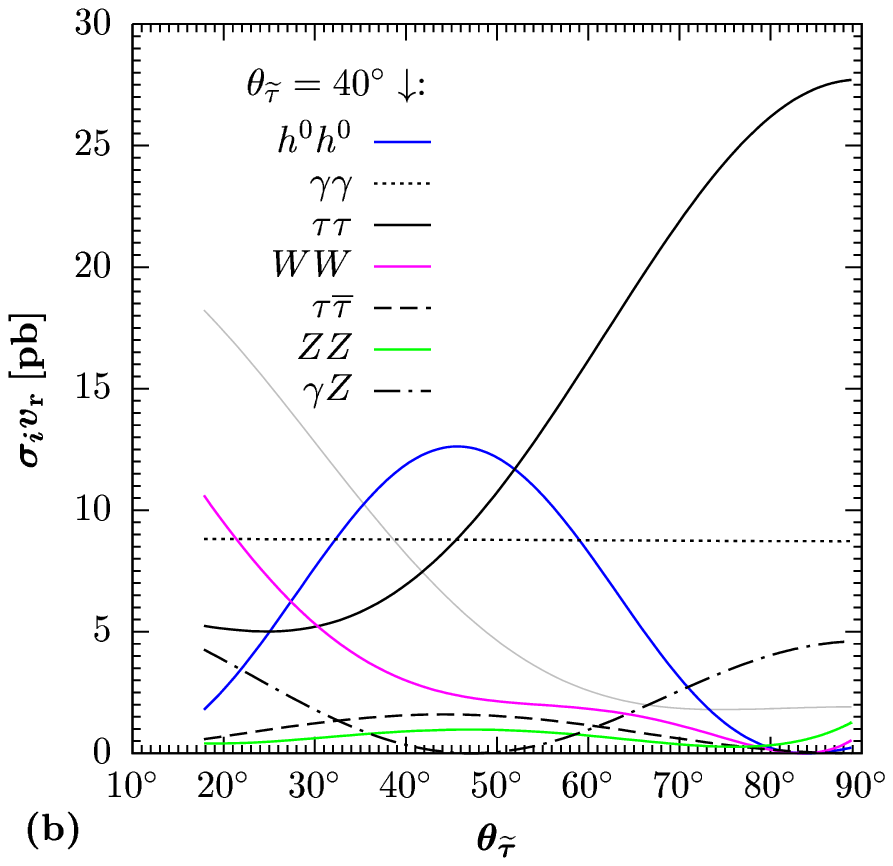}}
\caption{\small (a)~The dependence of $\mstautwo$ (curved solid line)
  and $m_{\widetilde{\nu}_{\tau}}$ (dashed line) on the stau mixing
  angle $\thetastau$ for the input parameters $\mstauone=130\ \GeV$
  (horizontal solid line), $\tanb=10$, $\mA=M_S=M_{3}=-A=1\ \TeV$, and
  $6M_1=M_2=\mu=1\ \TeV$ (i.e., $\neutralino\simeq\widetilde{B}$) for
  which $m_{\neutralino}=169\ \GeV$ (dash-dotted line) and
  $m_{\hhiggs}=116\ \GeV$ (dotted line). (b)~Dominant stau
  annihilation cross sections times the relative velocity
  $v_{\mathrm{r}}$ of the incoming staus as a function of $\thetastau$
  for $\pcm=10\ \GeV$ and the same input parameters as in~(a). The
  curves show the channels with the following final states:
  $\hhiggs\hhiggs$, $\gamma\gamma$, $\tau\tau$, $WW$,
  $\tau\overline{\tau}$, $ZZ$, $\gamma Z$ (at $\thetastau=40^\circ$,
  from top to bottom).  In addition, we plot
  $\sigma_{\tau\tau}v_{\mathrm{r}}$ for the case of a wino-like
  neutralino, $\neutralino\simeq\widetilde{W}$, with
  $m_{\neutralino}=175\ \GeV$ as obtained with $M_{1}=6M_{2}=1\ \TeV$
  (thin gray line). No lines are shown for $\thetastau < 18^\circ$
  where $m_{\widetilde{\nu}_{\tau}}<\mstauone$.}
\label{Fig:mix}
\end{center}
\end{figure}

In Fig.~\ref{Fig:mix}a we show the $\thetastau$-dependence of the
masses of the heavier stau, $m_{\stautwo}$, (curved solid line) and
the tau-sneutrino, $m_{\widetilde{\nu}_{\tau}}$, (dashed line) for
fixed $\mstauone = 130\ \GeV$ and the input parameters $\tanb=10$,
$\mA=\mu=-A=1\ \TeV$, and $6 M_{1}=M_{2,3}=1\ \TeV$.  Because of SU(2)
gauge invariance, $\mstauL$ sets also the soft-breaking mass for the
tau-sneutrino hence approximately $m^2_{\widetilde{\nu}_{\tau}} \sim
\mstauR^2 + \delta$ so that $\widetilde{\nu}_{\tau}$ becomes lighter
than \stauone\ for $\thetastau \lesssim 18^\circ$ ($\delta$ is
negative in that region).  In addition, we plot the masses of the
lightest neutralino, $m_{\neutralino}=169\ \GeV$ (dash-dotted line),
the lighter stau, $\mstauone = 130\ \GeV$ (horizontal solid line), and
the lightest Higgs, $m_{\hhiggs}=116\ \GeV$ (dotted line). We note in
passing that $\mstauone$ may deviate slightly from its anticipated
input value due to radiative corrections. We then correct for this by
an adjustment of $\mstauR^2$ so that we indeed ensure $\mstauone$ to
be constant.

In Fig.~\ref{Fig:mix}b we plot the dominant stau annihilation cross
sections times the relative (non-relativistic) velocity in the
center-of-mass frame of the incoming staus, $v_{\mathrm{r}} =
2\pcm/\mstauone$, for the same parameters as in Fig.~\ref{Fig:mix}a.
Owing to an (approximate) Maxwell-Boltzmann distribution of the stau
velocity, $\langle \pcm \rangle|_{\Tf} \sim \sqrt{\mstauone\Tf}$, we
choose $\pcm = 10\ \GeV$ as a representative value.%
\footnote{This value is actually at the somewhat lower end, given
  $\mstauone\gtrsim 100\ \GeV$ and $\Tf\simeq \mstauone/25$. However,
  $\sigma v_{\mathrm{r}}$ depends only weakly on $\pcm$, and the
  thermally averaged $\sigmavof{i}$ will be shown in the upcoming
  figures.}
The curves show the annihilation channels with the following final
states: $\hhiggs\hhiggs$, $\gamma\gamma$, $\tau\tau$, $WW$,
$\tau\overline{\tau}$, $ZZ$, $\gamma Z$ (at $\thetastau=40^\circ$,
from top to bottom).
All channels except $\gamma\gamma$ show a strong dependence on
\thetastau. The $\hhiggs\hhiggs$ ($\tau\overline{\tau}$) channel peaks
at $\thetastau=\pi/4$---a feature which we will discuss in detail in
Sect.~\ref{sec:enhanc-coupl-higgs}. For the $\tau\tau$ channel, the
overall size of the cross section is governed by $m_{\neutralino}$
since this channel proceeds only via $t(u)$-channel exchanges of
neutralinos. Our chosen input values lead to a bino-like neutralino,
$\neutralino\simeq\widetilde{B}$, and $\sigma_{\tau\tau}$ drops for an
increasingly `left-handed' stau. (For comparison, the thin gray line
shows $\sigma_{\tau\tau}v_{\mathrm{r}}$ for the case of a wino-like
lightest neutralino, $\neutralino=\widetilde{W}$, of similar mass,
$m_{\neutralino}=175\ \GeV$, as obtained by changing the gaugino mass
input parameters to $M_{1}=6M_2=1\ \TeV$.)  The annihilation into a
$WW$ pair becomes important for an increasing $\stauL$ component in
$\stauone$, i.e., towards smaller $\thetastau$, since the
$t(u)$-channel exchange with the tau-sneutrino opens up; the
$\stauone\widetilde{\nu}_{\tau}W$ ($\stauone\stauone WW$) coupling is
proportional to $\cos{\thetastau}$ ($\cos^2{\thetastau}$).  The
modulation of the $\gamma Z$ channel can be understood by considering
the structure of the $\stauone\stauone Z$ coupling $\propto (1 -
4\sin^2{\theta_W} + \cos{2\thetastau})$. Note that the first two terms
practically cancel out. For stau annihilation into a $ZZ$ pair there
is an additional contribution from $\stautwo$-exchange with the
respective $\stauone\stautwo Z$ coupling $\propto \sin{2\thetastau}$.
Having discussed the dominant $\stauone$ annihilation channels in a
simple manner, we also warn the reader that interferences between the
different Feynman diagrams of a given channel may well lead to a
counterintuitive behavior. In this regard, see
Ref.~\cite{Berger:2008ti} for a thorough discussion of
$\stauone\stauone^*$ annihilation into vector bosons. For the limiting
case of a purely `right-handed' stau, $\stauone\simeq\stauR$
($\thetastau\rightarrow \pi/2$), we recover the relative importance of
the annihilation cross sections into $\gamma\gamma$, $\gamma Z$, $Z
Z$, and $\tau\tau$ with bino $t(u)$-channel exchange found in
Ref.~\cite{Asaka:2000zh}.

\begin{figure}[t]
\begin{center}
\includegraphics[width=0.65\textwidth]{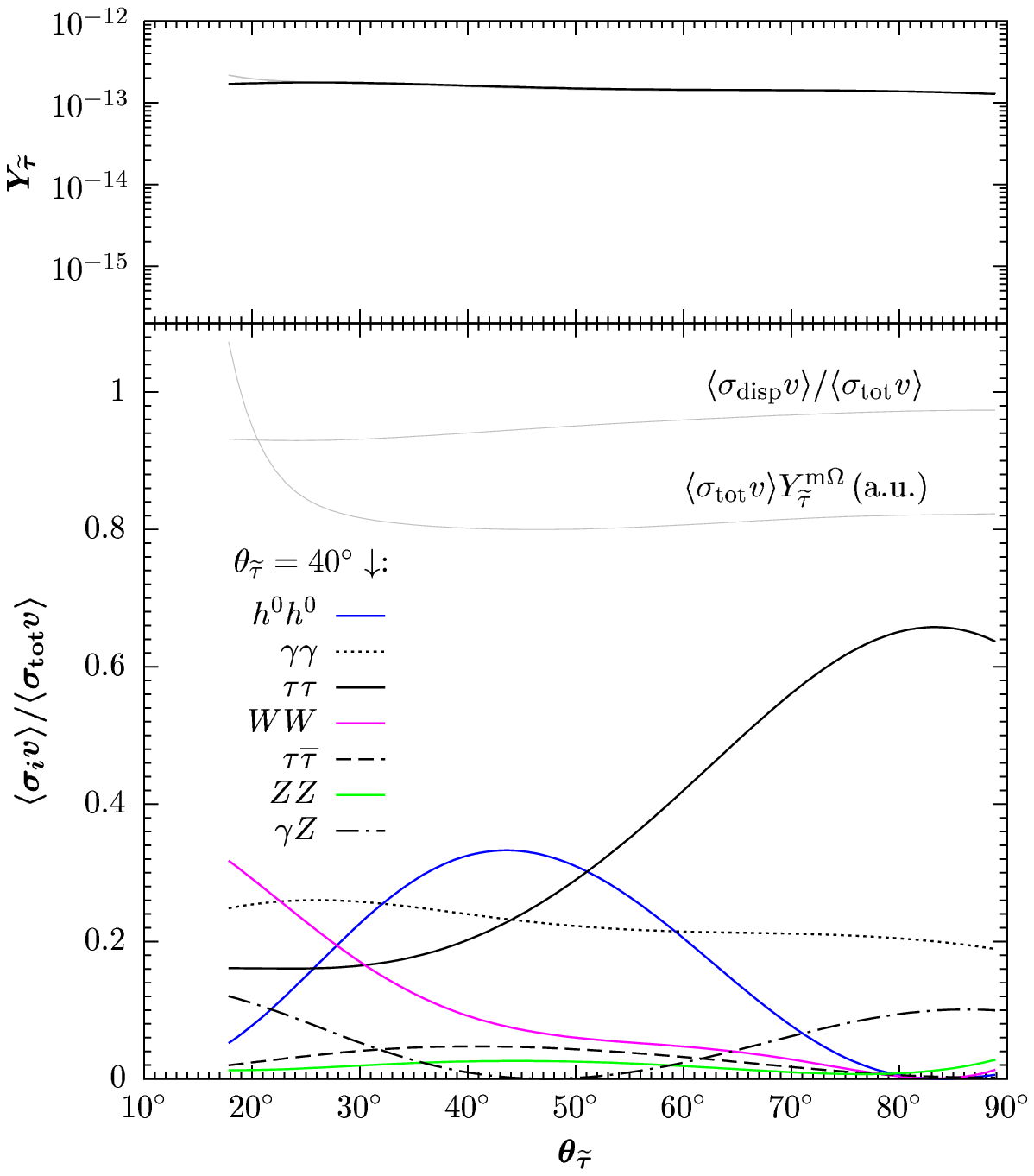}%
\caption{\small Dependence of the stau yield $\Ystau$ (upper panel)
  and of the relative importance of the dominant thermally averaged
  cross sections, $\sigmavof{i}/\sigmavof{\mathrm{tot}}$, at $x=25$
  (lower panel) on the stau-mixing angle $\thetastau$ for the same
  input parameters as in Fig.~\ref{Fig:mix}.  In the upper panel, the
  thick line shows the stau yield $\Ystau$ obtained with our relic
  abundance calculation and the thin gray line the one obtained with
  \texttt{micrOMEGAs} to which we refer as $\YstaumO$. In the lower
  panel, the line styles are associated with the same dominant
  annihilation channels as in Fig.~\ref{Fig:mix}b.  In addition, we
  show (as labeled) the relative importance of the sum of the
  displayed cross sections,
  $\sigmavof{\mathrm{disp}}/\sigmavof{\mathrm{tot}}$, and
  $\sigmavof{\mathrm{tot}}\,\YstaumO$ in arbitrary units (a.u.). No
  lines are shown for $\thetastau < 18^\circ$ where
  $m_{\widetilde{\nu}_{\tau}}<\mstauone$.}
\label{Fig:multi1}
\end{center}
\end{figure}

Figure~\ref{Fig:multi1} shows the $\thetastau$-dependence of $\Ystau$
(upper panel) and of the relative importance of the dominant
\textit{thermally averaged} cross sections,
$\sigmavof{i}/\sigmavof{\mathrm{tot}}$, at $x=25$ (lower panel) for
the same input parameters as in Fig.~\ref{Fig:mix}.
The lines in the lower panel are associated with the same dominant
annihilation channels as in Fig.~\ref{Fig:mix}b.  In addition, the
relative importance of the sum of the displayed cross sections,
$\sigmavof{\mathrm{disp}}/\sigmavof{\mathrm{tot}}$, (thin line, as
labeled) is shown to demonstrate that the displayed channels
constitute indeed (up to at most about $10\%$) the dominant part of
$\sigmavof{\mathrm{tot}}$ for the chosen set of input parameters.
In the upper panel, the total stau decoupling yield obtained with our
own relic abundance calculation is shown by the thick line and the one
computed with \texttt{micrOMEGAs}, $\YstaumO$, by the thin gray line.
For $\thetastau\lesssim 25^\circ$, both curves start to deviate from
each other since one enters the $\widetilde{\nu}_{\tau}$--$\stauone$
coannihilation region in which the stau decoupling yield increases.
This coannihilation effect leads also to the rise of the thin gray
line that shows $\langle\sigma_{\mathrm{tot}}v\rangle\,\YstaumO$ in
arbitrary units (a.u.) in the lower panel. Note that the same line
illustrates $\Ystau\propto 1/\sigmavof{\mathrm{tot}}$ for
$\thetastau>25^\circ$, where the result of our relic abundance
calculation agrees with $\YstaumO$.  Interestingly, for the given
input parameters, $\Ystau$ is not overly affected by the variation in
$\thetastau$ in this region, which reflects the fact that
$\sigmavof{\mathrm{tot}}$ and thereby $\sigmav$ vary by less than a
factor of about $1.5$ at the relevant time of decoupling.
In the next sections, we will demonstrate that this picture changes
significantly for certain other choices of the input parameters.

\section{Effects of large stau-Higgs couplings}
\label{sec:enhanc-coupl-higgs}

Owing to the scalar nature of the stau, there exists a remarkable
difference between the standard neutralino decoupling and the scenario
in which the long-lived stau freezes out from the primordial plasma.
For the neutralino LSP, the $\mu$ parameter enters into the
annihilation cross sections only indirectly by influencing the
gaugino/higgsino mixture of $\neutralino$.  This stands in strong
contrast to the case in which a scalar particle is the lightest
Standard Model superpartner: the sfermions couple directly to
dimensionful parameters of the theory, namely, the trilinear couplings
$A$ and the Higgs-higgsino mass parameter $\mu$.  The corresponding
operators in the MSSM Lagrangian always contain a Higgs field. In
particular, the stau--Higgs couplings are given by
\begin{equation}
    \label{eq:L-stau-stau-Higgses}
    \Lagrangian_{\mathrm{MSSM}} \ni\frac{g}{\mW} \sum_{\alpha , \beta = \L , \R} \stau^*_{\alpha} 
    \couptriLR{\stau^*_{\alpha}}{\stau_{\beta}}{ \mathcal{H}}
    \stau_{\beta} \mathcal{H} 
\end{equation}
with $ \mathcal{H} = \hhiggs,\,\Hhiggs ,\,\Ahiggs$.  We have pulled
out the factor $g/\mW$ so that the `reduced' couplings
$\couptriLR{\stau^*_{\alpha}}{\stau_{\beta}}{ \mathcal{H}}$ among the
gauge eigenstates $\stauL$ and $\stauR$ are given
by~\cite{Haber:1997dt}
\begin{align}
    \label{eq:lighthiggs-stau-stau-couplings}
     \couptriLR{\stau^*}{\stau}{\hhiggs} & = 
    \begin{pmatrix}
      \displaystyle
      \left( -{\frac{1}{2}} + \ssqw \right) \mZ^2 \sapb + \mtau^2
      {\frac{\sa}{\cb}} 
      &
      \displaystyle
      \frac{\mtau}{2} \left( \Atau \frac{\sa}{\cb}
        + \mu \frac{\ca}{\cb} \right) 
      \\ 
      \displaystyle
      \frac{\mtau}{2} \left( \Atau
        \frac{\sa}{\cb} + \mu \frac{\ca}{\cb} \right) 
      & 
      \displaystyle
      - \ssqw \mZ^2
      \sapb + \mtau^2 {\frac{\sa}{\cb}}
    \end{pmatrix}\ ,
\\
    \label{eq:CPoddhiggs-stau-stau-couplings}
     \couptriLR{\stau^*}{\stau}{\Ahiggs}  &=
     \begin{pmatrix}
      0
      &
      \displaystyle
      +i \frac{\mtau}{2} \left(  \Atau \tanb + \mu\right) 
      \\ 
      \displaystyle
      -i \frac{\mtau}{2} \left(  \Atau \tanb + \mu\right) 
      & 
      0
    \end{pmatrix}\ ,
\end{align}
where $\couptriLR{\stau^*}{\stau}{\Hhiggs}$ can be obtained from
(\ref{eq:lighthiggs-stau-stau-couplings}) upon the replacement $\alpha
\rightarrow \alpha-\pi/2$.  Whenever convenient, we use the shorthand
notation $\ssqw=\sin^2{\theta_W}$, $c_{\gamma}=\cos{\gamma}$, and
$s_{\gamma}=\sin{\gamma}$. The parameters $\Atau$ and $\mu$ only
appear off-diagonal and they are multiplied with the associated
fermion mass, the tau mass $\mtau$.

Using $C = \stauROT \widetilde{C} \stauROT^\dagger$, one obtains the
couplings of the mass eigenstates $\stauone$ and $\stautwo$. In this
regard, it is important to note that the coupling of the CP-odd Higgs
boson to the lighter stau vanishes,
$\couptri{\stau^*_{1}}{\stau_{1}}{\Ahiggs}=0$. Therefore, we have not
listed the process $\stauone\stauone^*\rightarrow\gamma\Ahiggs$ in
Table~\ref{tab:ann-channels}. By the same token, there is also no
$s$-channel exchange of $\Ahiggs$ in any of the annihilation channels.
Note that this statement remains valid even after the inclusion of
radiative corrections: There is no induced mixing between
$\hhiggs(\Hhiggs)$ and $\Ahiggs$ in absence of CP-violating effects in
the SUSY sector.

Let us now turn to the probably most interesting couplings in the
context of $\stauone\stauone^*$ annihilation, namely, the ones of the
lighter stau to $\hhiggs$ and $\Hhiggs$. The `reduced'
$\stauone\stauone\hhiggs$ coupling reads
  \begin{align}
    \label{eq:higgses-stau-stau-couplings-PHYS}
     \couptri{\stau^*_{1}}{\stau_{1}}{\hhiggs}& =
      \displaystyle
      \left( -{\frac{1}{2}} \csqth + \ssqw \ctwoth \right) \mZ^2 \sapb + \mtau^2
      {\frac{\sa}{\cb}} + 
       \frac{\mtau}{2} \left( \Atau \frac{\sa}{\cb}
        + \mu \frac{\ca}{\cb} \right) \stwoth \ .
\end{align}
This is a complicated expression. However, if we choose $\mA$ to be
large, $\mA\gg\mZ$, we can
simplify~(\ref{eq:higgses-stau-stau-couplings-PHYS}) by using
$\cos{(\beta-\alpha)}=0$ [cf.~(\ref{eq:DL-cbma})],
\begin{align}
  \label{eq:coup-s1-s1-h-DL}
  \couptriDL{\stau^*_{1}}{\stau_{1}}{\hhiggs} & \simeq \displaystyle \left(
    {\frac{1}{2}} \csqth - \ssqw \ctwoth \right) \mZ^2 \ctwob -
  \mtau^2 - \frac{\mtau}{2} \Xtau \stwoth \ .
\end{align}
Thereby, we make an interesting observation: In the decoupling limit
(DL), the $\stauone\stauone\hhiggs$ coupling becomes proportional to
the left--right entry $\mtau \Xtau$ of the stau mass-squared matrix
(\ref{eq:stau-mass-matrix}) and to $\stwoth$. Therefore, it comes as
no surprise that the $\stauone\stauone^*$ annihilation cross section
into $\hhiggs\hhiggs$ peaks at $\thetastau = \pi/4$---the point of
maximal $\stauL$-$\stauR$ mixing---as can be seen, e.g., in
Fig.~\ref{Fig:mix}b. Analogously, one finds that the
$\stauone\stauone\Hhiggs$ coupling is proportional to
$\left(\Atau\tanb + \mu\right) \stwoth$ in the decoupling limit.
Complementary, the $\stauone\stautwo\hhiggs/\Hhiggs$ couplings exhibit
in this limit the same combination of $A$, $\mu$, and $\tanb$ as their
$\stauone\stauone$ counterparts but those terms are now multiplied by
$\ctwoth$ instead of $\stwoth$.

After the above discussion, it is clear that there exists the
possibility to enhance the total stau annihilation cross section
$\sigmatot$---and thereby to decrease $\Ystau\propto
1/\sigmavof{\mathrm{tot}}$---by choosing a proper combination of large
$A$, $\mu$, and $\tanb$. In the remainder of this section, we will
explore this possibility for two exemplary pMSSM scenarios.

Before proceeding let us make some technical comments. Large values of
the previously mentioned parameters may well lead to large radiative
corrections.%
\footnote{In this context, note that we introduce a large
  $m_\mathrm{t}$--$m_{\widetilde{\mathrm{t}}_{1,2}}$ splitting when
  choosing $M_S= 1\ \TeV$.}
In order to arrive at a proper loop-improved tree-level result, we
re-evaluate the entire Higgs sector using \texttt{FeynHiggs}.  In
particular, we have modified our generated matrix elements in a way
that allows us to set all trilinear Higgs couplings to their
loop-corrected values.%
\footnote{We are grateful to T.~Plehn and M.~Rauch for providing us,
  for cross-checking, with their implementation of a \texttt{Fortran}
  routine which calculates the Higgs self-couplings using the
  effective potential approach~\cite{Carena:1995bx}.}
Note that this goes well beyond a simple
$\alpha\rightarrow\alpha_{\mathrm{eff}}$ prescription. Only then, we
mostly find better agreement of our cross sections for stau
annihilation into two Higgses with the ones computed by
\texttt{micrOMEGAs}.  The latter program uses \texttt{CalcHEP}
\cite{Pukhov:2004ca} for the generation of the matrix elements. There,
the trilinear Higgs self-couplings have been expressed in terms of
$m_{\hhiggs}$, $m_{\Hhiggs}$, and $m_{\Ahiggs}$ which effectively
reabsorbs a bulk of the radiative corrections~\cite{Dubinin:1998nt}.
We therefore think that we do slightly better whenever we encounter
some disagreement between the mentioned cross sections. Though the
overall effect on $\Ystau$ is typically small, it can be at the level
of $20\%$ (see below).  Finally, it is well known that a large $A$
parameter may lead to charge/color breaking minima (CCB) in the scalar
MSSM potential; see, e.g., Ref.~\cite{Casas:1995pd}. \texttt{SuSpect}
performs some basic checks which we take into account to make sure
that we do not violate the constraints associated with CCB. In any
case, our pMSSM scenarios shall be regarded as toy models which allow
us to extract important features of primordial stau annihilation in
the most transparent way.

\begin{figure}[t]
\begin{center}
\includegraphics[width=0.65\textwidth]{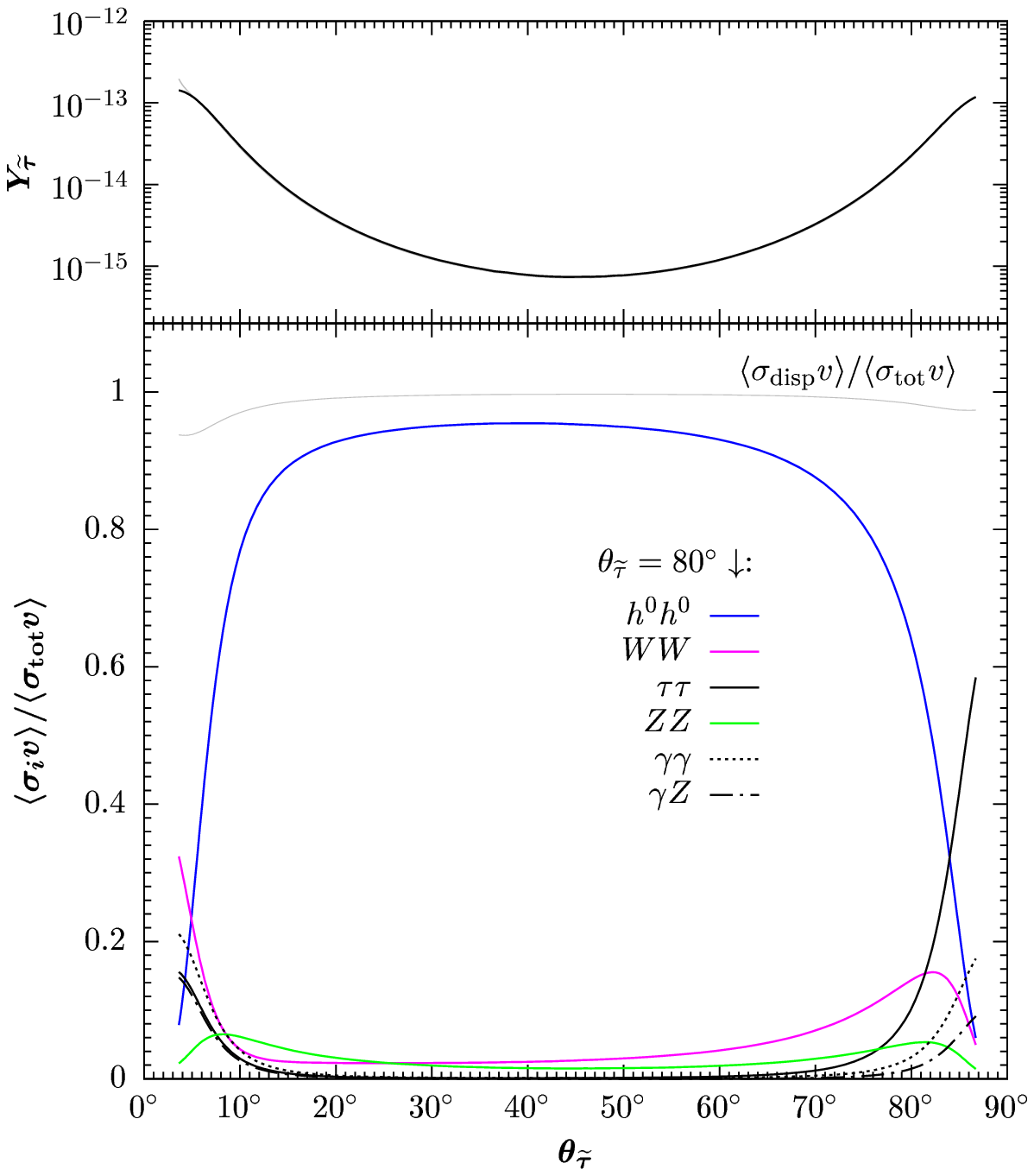} 
\caption{\small Analogous to Fig.~\ref{Fig:multi1} but for the pMSSM
  scenario associated with $\mstauone=130\ \GeV$, $\tanb=50$, and
  $\mA=\mu=M_S=6M_1=M_{2,3}=-A=1\ \TeV$ and for $x=30$. The stau
  decoupling yield takes on its minimum value of $\Ystau = 7.4\times
  10^{-16}$ at $\thetastau=45^\circ$. The displayed stau annihilation
  channels are associated with the following final states:
  $\hhiggs\hhiggs$, $WW$, $\tau\tau$, $ZZ$, $\gamma\gamma$, and
  $\gamma Z$ (at $\thetastau=80^\circ$, from top to bottom). No lines
  are shown for $\thetastau < 4^\circ$ where
  $m_{\widetilde{\nu}_{\tau}}<\mstauone$.}
\label{Fig:multi2}
\end{center}
\end{figure}

In Fig.~\ref{Fig:multi2} we demonstrate the effect associated with a
large $\stauone\stauone\hhiggs$ coupling by presenting the
$\thetastau$-dependence of $\Ystau$ (upper panel) and of the relative
importance of the dominant \textit{thermally averaged} cross sections,
$\sigmavof{i}/\sigmavof{\mathrm{tot}}$, at $x=30$ (lower panel) for
the pMSSM scenario associated with $\mstauone = 130\ \GeV$,
$\tanb=50$, $\mA=M_S=M_{3}=-A=1\ \TeV$, and $6M_1=M_2=\mu=1\ \TeV$.
In this scenario, $m_{\hhiggs}$ stays in the range $117-119\ \GeV$ and
the lightest neutralino is bino-like with a mass of
$m_{\neutralino}=169\ \GeV$.  Stau annihilation into heavy Higgses
remains kinematically forbidden.  The curves in the lower panel are
associated with stau annihilation into $\hhiggs\hhiggs$, $WW$,
$\tau\tau$, $ZZ$, $\gamma\gamma$, and $\gamma Z$ (at
$\thetastau=80^\circ$, from top to bottom).  As is evident, the
annihilation into $\hhiggs\hhiggs$ is enhanced already well before
$\thetastau=\pi/4$. At the peak position,
$\sigma_{\hhiggs\hhiggs}v_{\mathrm{r}}\simeq 8.8\times 10^3\
\mathrm{pb}$ for $\pcm = 10\ \GeV$ (no thermal average), which is
still three orders of magnitude below the unitarity bound for
inelastic s-wave annihilation,
$\sigma_{\mathrm{u}}v_{\mathrm{r}}=8\pi/(\mstauone\pcm)$
\cite{Griest:1989wd,Berger:2008ti}. Also the cross sections for stau
annihilation into $WW$ and $ZZ$ are strongly enhanced towards
$\thetastau=\pi/4$ since the $s$-channel contribution of
$\stauone\stauone\rightarrow {\hhiggs}^*\rightarrow VV$ becomes very
important. At their respective peak positions,
$\sigma_{WW}v_{\mathrm{r}}\simeq 250\ \mathrm{pb}$ and
$\sigma_{ZZ}v_{\mathrm{r}}\simeq 130\ \mathrm{pb}$ for $\pcm = 10\
\GeV$. (Because of the dominance of the $\hhiggs\hhiggs$ channel, the
corresponding maxima do not show up in Fig.~\ref{Fig:multi2} where
$\sigmavof{i}/\sigmavof{\mathrm{tot}}$ is shown.) By the same token,
the cross sections of all (kinematically allowed) channels with a
fermion-antifermion final state (e.g.\ $\tau\overline{\tau}$)---which
are subdominant in the scenario considered in
Fig.~\ref{Fig:multi2}---experience an enhancement for
$\thetastau\rightarrow \pi/4$.  In total, there is an enhancement of
$\sigmavof{\mathrm{tot}}$ that delays the thermal freeze out of the
staus significantly, i.e., $x_{\mathrm f}\simeq 33$ for
$\thetastau\simeq\pi/4$.  As can be seen in the upper panel of
Fig.~\ref{Fig:multi2}, the decoupling yield is thereby reduced
dramatically down to a minimum value of $\Ystau=7.4\times 10^{-16}$
for maximal left--right mixing of the staus.

\begin{figure}[t]
\begin{center}
\includegraphics[width=0.7\textwidth]{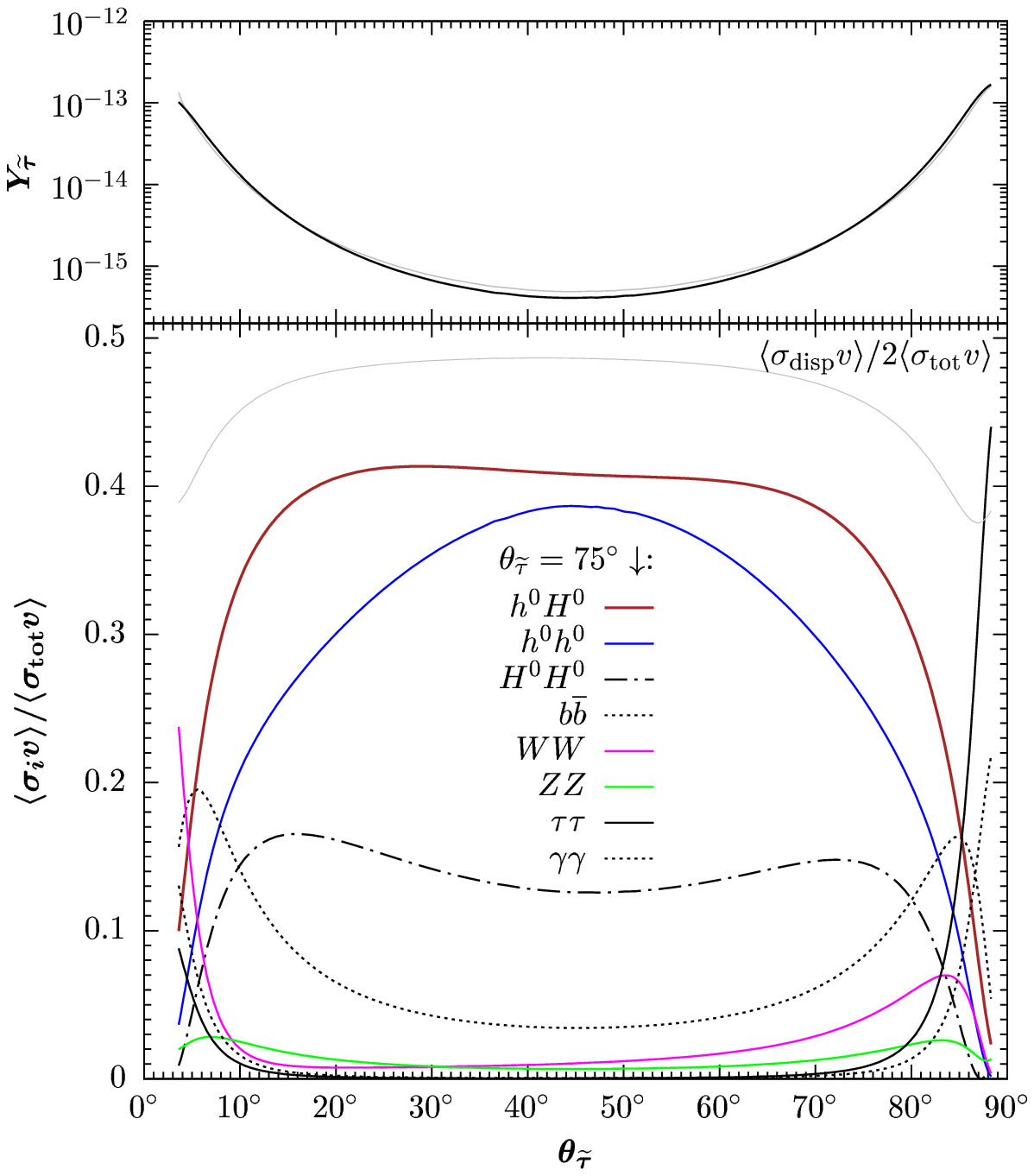} 
\caption{\small Analogous to Fig.~\ref{Fig:multi2} but for the pMSSM
  scenario associated with $\mstauone=150\ \GeV$, $\tanb=50$,
  $\mA=130\ \GeV$, $M_S=M_{3}=-A=1\ \TeV$, $3M_1=M_2=\mu=1\ \TeV$ and
  for $x=30$.  The stau decoupling yield reaches its minimum value of
  $\Ystau = 4.1\times 10^{-16}$ at $\thetastau=\pi/4$. The displayed
  stau annihilation channels are associated with the following final
  states: $\hhiggs\Hhiggs$, $\hhiggs\hhiggs$, $\Hhiggs\Hhiggs$,
  $b\overline{b}$, $WW$, $ZZ$, $\tau\tau$, and $\gamma\gamma$ (at
  $\thetastau=75^\circ$, from top to bottom). For an optimized
  presentation of those channels, the line indicating the relative
  importance of the sum of the displayed cross sections is scaled down
  by a factor of 1/2:
  $\sigmavof{\mathrm{disp}}/2\sigmavof{\mathrm{tot}}$.  No lines are
  shown for $\thetastau < 4^\circ$ where
  $m_{\widetilde{\nu}_{\tau}}<\mstauone$.}
\label{Fig:low-mA}
\end{center}
\end{figure}

In the previous pMSSM examples, annihilation into final states
containing heavy Higgs bosons is kinematically forbidden. We can allow
for those channels by reducing the input value $\mA$. Indeed,
scenarios in which all Higgs bosons are very light in conjunction with
large $\tanb$ have been studied in the literature, see,
e.g.,~\cite{Boos:2002ze,Boos:2003jt} and references therein.  We thus
consider now the following pMSSM scenario: $\mA=130\ \GeV$,
$\mstauone=150\ \GeV$, $\tanb=50$, $M_S=M_{3}=-A=1\ \TeV$, and
$3M_1=M_2=\mu=1\ \TeV$.  In Fig.~\ref{Fig:low-mA}, the associated
$\thetastau$-dependence of $\Ystau$ and of
$\sigmavof{i}/\sigmavof{\mathrm{tot}}$ at $x=30$ for the now dominant
channels is shown in a similar way as in Fig.~\ref{Fig:multi2}; only
the relative importance of the sum of the displayed cross sections is
scaled down by a factor of 1/2,
$\sigmavof{\mathrm{disp}}/2\sigmavof{\mathrm{tot}}$, to allow for an
optimized presentation of the single dominant channels.  Throughout
the considered $\thetastau$ range, the masses of both CP-even Higgs
bosons are relatively light and remain rather constant:
$\mh=(118\pm1.5)\ \GeV$ and $\mH=(128.5\pm1)\ \GeV$.  Here the
dominant annihilation channels are associated with the following final
states: $\hhiggs\Hhiggs$, $\hhiggs\hhiggs$, $\Hhiggs\Hhiggs$,
$b\overline{b}$, $WW$, $ZZ$, $\tau\tau$, and $\gamma\gamma$ (at
$\thetastau = 75^\circ$, from top to bottom).  As can be seen, stau
annihilation into $\hhiggs\Hhiggs$ is now more dominant than the one
into $\hhiggs\hhiggs$ and also the $\Hhiggs\Hhiggs$ channel becomes
important, where each of those channels is indeed associated with an
(absolute) annihilation cross section $\sigmavof{i}$ that peaks at
$\thetastau=\pi/4$.  Also the annihilation into $b\overline{b}$ is
significant---a process which we will discuss in detail in the
following section. In this respect, one should stress that all
processes with $s$-channel \Hhiggs\ exchange are here less suppressed
by $\mH^2$ in the respective propagator than in the previously
considered scenarios.  Note that the asymmetry of
$\sigmavof{i}/\sigmavof{\mathrm{tot}}$ of those dominant channels
($\hhiggs\Hhiggs$, $\hhiggs\hhiggs$, $\Hhiggs\Hhiggs$,
$b\overline{b}$) with respect to a reflection at $\thetastau=\pi/4$ is
dominantly caused by the $\thetastau$-dependent modulation of the $WW$
channel.  As in the pMSSM scenario considered in
Fig.~\ref{Fig:multi2}, there is again an significant enhancement of
$\sigmavof{\mathrm{tot}}$ that delays the stau freeze out such that
$x_{\mathrm f}\simeq 33$ at $\thetastau\simeq\pi/4$.  Thereby, the
efficient annihilation into final state Higgses is accompanied by a
significant drop in \Ystau\ down to $\Ystau= 4.1\times 10^{-16}$ at
$\thetastau=\pi/4$ as can be seen in Fig.~\ref{Fig:low-mA}. At this
minimum, there is a $20\%$ disagreement between $\Ystau$ from our
calculation of stau decoupling (solid line) and the
\texttt{micrOMEGAs} result $\YstaumO$ (thin gray line) which is a
consequence of the different treatments of the Higgs sector described
above.

Let us finally remark that the Higgs couplings to fermions and vector
bosons as well as the Higgs self-couplings develop a strong dependence
on $\mA$ once we leave the decoupling regime ($\mA\lesssim 200\
\GeV$); for a comprehensive review see, e.g.,
Ref.~\cite{Djouadi:2005gj}.%
\footnote{The Higgs sector is also particularly sensitive to the
  mixing in the stop sector. In the considered pMSSM scenarios,
  $|X_t|\equiv |A_t-\mu\cot{\beta}|\sim M_S$ which corresponds to the
  ``typical-mixing scenario''~\cite{Carena:1999xa}.}
Changes in $\mA$ can therefore be accompanied by shifts in the
relative importance of the corresponding annihilation cross sections.
This underlines the fact that the details in the Higgs sector may very
well be crucial for the determination of the relic abundance of a
long-lived $\stauone$.

\section{Resonant stau annihilation}
\label{sec:reson-annih}

\begin{figure}[t]
\begin{center}
\includegraphics[width=0.7\textwidth]{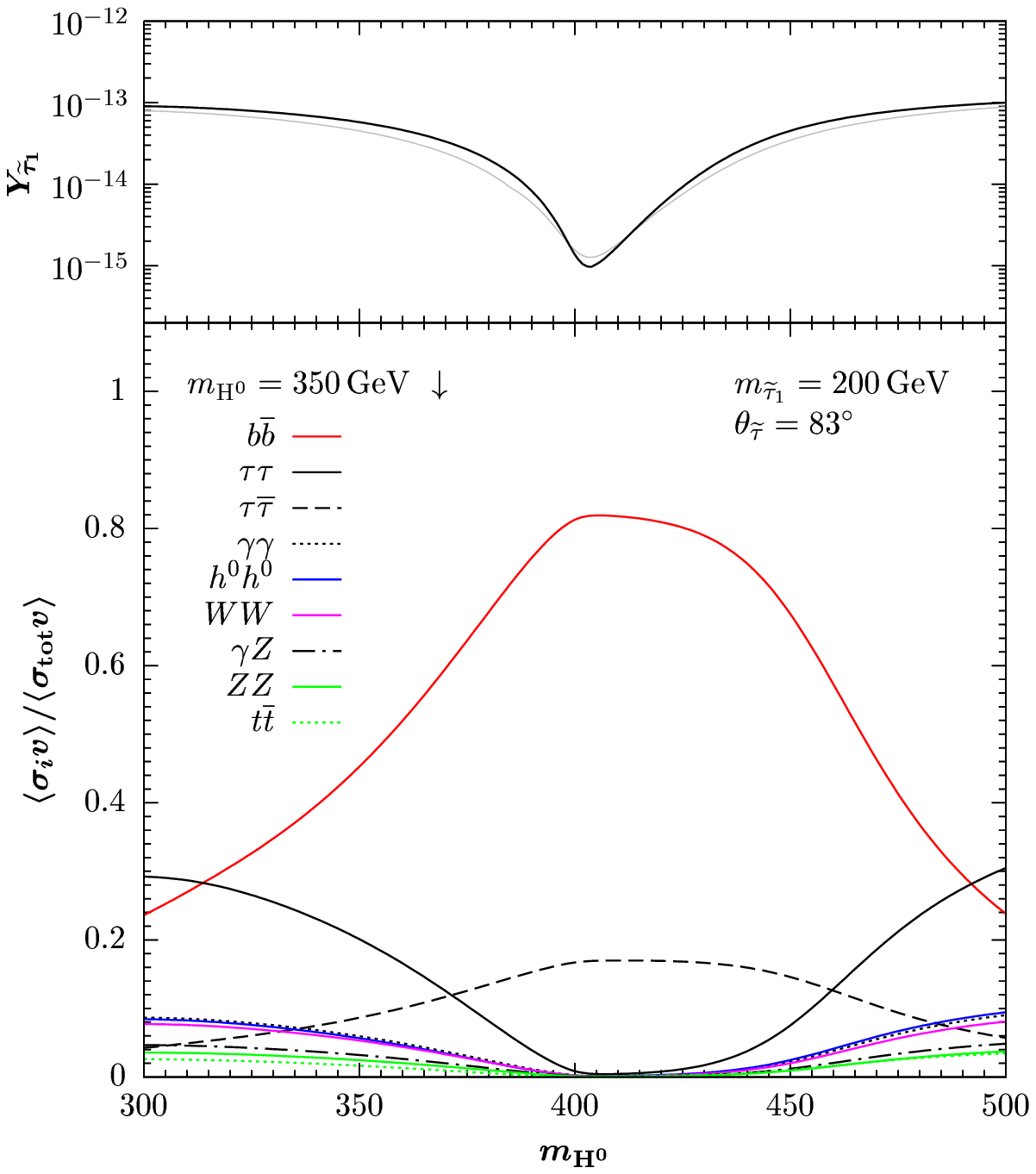} 
\caption{\small Dependence of $\Ystau$ (upper panel) and of
  $\sigmavof{i}/\sigmavof{\mathrm{tot}}$ at $x=25$ (lower panel) on
  $\mH$ for the pMSSM scenario associated with $\mstauone=200\ \GeV$,
  $\thetastau=83^\circ$, $\tanb=40$, and $-A=\mu=4M_1=M_{2,3}=M_S=1\
  \TeV$. In the upper panel, the dark line shows the stau yield
  $\Ystau$ obtained with our relic abundance calculation and the thin
  gray line the one obtained with \texttt{micrOMEGAs}. The stau
  decoupling yield takes on its minimum value of $\Ystau = 9.7\times
  10^{-16}$ at $\mH=404~\GeV$. In the lower panel, the displayed
  dominant stau annihilation channels are associated with the
  following final states: $b\overline{b}$, $\tau\tau$,
  $\tau\overline{\tau}$, $\gamma\gamma$, $\hhiggs\hhiggs$, $WW$,
  $\gamma Z$, $ZZ$, and $t\overline{t}$ (at $\mH=350\ \GeV$, from top
  to bottom).}
\label{Fig:res}
\end{center}
\end{figure}

By inspection of Table~\ref{tab:ann-channels} it becomes clear that
primordial stau annihilation can also proceed resonantly via
$s$-channel exchange of the heavy CP-even Higgs boson~$\Hhiggs$ for
$\mH\simeq 2\mstauone$. While the LEP bound on the stau mass
$\mstauone \gtrsim 82\ \GeV$~\cite{Yao:2006px} forbids \hhiggs\ to
become on-shell ($\mh^{\mathrm{max}} \sim 140~\GeV$, e.g.,
\cite{Djouadi:2005gj}), the $s$-channel exchange of \Ahiggs\ is
absent%
\footnote{Even in absence of SUSY-induced CP violation, resonant
  annihilation via $\Ahiggs$-exchange may still proceed through
  $\stauone$-$\stautwo$ coannihilation. However, this scenario
  requires considerable fine-tuning in the stau mass-squared matrix
  since $\stauone$ and $\stautwo$ have to be nearly degenerate.}
because of \mbox{$C[\stauone,\stauone^*,\Ahiggs] = 0$} (see
Sect.~\ref{sec:enhanc-coupl-higgs}).  Again, our choice to work in the
framework of the pMSSM proves to be very helpful. Since the $\Hhiggs$
resonance occurs for $2\mstauone\simeq m_{\Hhiggs}$, one runs quickly
into the decoupling limit in which $\mH$ is governed by the input
parameter $\mA$ according to the simple
relation~(\ref{eq:DL-higgs-masses}). This allows us to scan through
the resonance easily.

Let us explore resonant stau annihilation by considering the exemplary
pMSSM scenario associated with $\mstauone=200\ \GeV$,
$\thetastau=83^\circ$ (i.e., a mostly `right-handed' $\stauone$),
$\tanb=40$, and $-A=\mu=4M_1=M_{2,3}=M_S=1\ \TeV$, for which we vary
$\mA$ (and thereby $\mH$) to scan through the resonance.
Figure~\ref{Fig:res} shows the resulting $\mH$-dependence of $\Ystau$
(upper panel) and of $\sigmavof{i}/\sigmavof{\mathrm{tot}}$ at $x=25$
for the dominant annihilation channels (lower panel). Those channels
are now associated with the following final states: $b\overline{b}$,
$\tau\tau$, $\tau\overline{\tau}$, $\gamma\gamma$, $\hhiggs\hhiggs$,
$WW$, $\gamma Z$, $ZZ$, and $t\overline{t}$ (at $\mH=350\ \GeV$, from
top to bottom).  In Table~\ref{tab:ann-channels} all resonant channels
can be identified.  Close to the resonance condition $2\mstauone\simeq
m_{\Hhiggs}$, the most important processes are stau annihilation into
$b\overline{b}$ and $\tau\overline{\tau}$. This is because the
couplings of those final state fermions to $\Hhiggs$ are $\tanb$
enhanced: for $\tanb\gg 1$, the \mbox{$f$$\overline{f}$$\Hhiggs$}
coupling $\sim m_{f} \sbma \tanb$ with $f=b,\tau$~\cite{Haber:1997dt}.
The (broad) peak associated with the resonance%
\footnote{Notice that we plot $\sigmavof{i}/\sigmavof{\mathrm{tot}}$
  so that the actual shape of the resonance looks somewhat different.}
already builds up for $\mH>2\mstauone=400\ \GeV$.  At zero relative
velocity, this would be a region in which the $\Hhiggs$ resonance
cannot occur. However, since $\stauone$ is in kinetic equilibrium at
the time of freeze out, resonant annihilation takes place already for
$2\mstauone<\mH$~\cite{Griest:1990kh}. For $\mH<2\mstauone=400\ \GeV$,
the processes containing $s$-channel $\Hhiggs$ exchange proceed with a
slightly faster rate (if kinematically allowed).  The impact of the
$\Hhiggs$ resonance on the thermal $\stauone$ freeze out and the
resulting $\Ystau$ is substantial.  Since the total width of $\Hhiggs$
is $\Gamma_{\Hhiggs} = (6-10)\ \GeV$ for $m_{\Hhiggs}= (300-500)\
\GeV$ in the considered pMSSM scenario, the reduction of $\Ystau$
extends over a relatively large $\mH$ range. In this regard, note that
$\Gamma_{\Hhiggs} $ could be substantially larger had we not
essentially decoupled all sfermions---except $\stauone$, $\stautwo$,
and $\widetilde{\nu}_{\tau}$---by choosing $M_S=1\ \TeV$.  For
$\mH\simeq 404~\GeV$, i.e., at the dip of the resonance, we find
$x_{\mathrm f}\simeq 33$ and a minimum stau decoupling yield of
$\Ystau=9.7\times 10^{-16}$ (dark line).  Thus, despite the (still)
moderate value of $\tanb=40$, a significant reduction of $\Ystau$ is
encountered. Indeed, $\Ystau$ can be even further suppressed for a
larger value of $\tanb$.  Let us remark that an accurate determination
of $\Ystau$ in the resonance region requires to take special care of
the $b\overline{b}\Hhiggs$ vertex.  This coupling is well known to
receive substantial radiative corrections for sizable values of
$\tanb$.  Therefore, we rely again on the computer tool
\texttt{FeynHiggs} to compute all quark--antiquark--Higgs couplings
and the total width $\Gamma_{\Hhiggs}$.  Also the \texttt{micrOMEGAs}
code takes special care of the $b\overline{b}\Hhiggs$ vertex.  We
therefore think that the difference between the yields shown in the
upper panel of Fig.~\ref{Fig:res} reflects the theoretical uncertainty
involved in the determination of $\Gamma_{\Hhiggs}$ as well as the
$b\overline{b}\Hhiggs$ vertex.

\section{On the viability of a \boldmath$\stauone$-$\stauone^*$ asymmetry}
\label{sec:comm-stau-stau}

Given the strong bounds on the abundance of negatively charged
$\stauone$ from bound-state effects during BBN, i.e., from CBBN of
$^6$Li and $^9$Be, it is natural to ask whether it is possible to have
an excess of positively charged $\stauone^*$'s over negatively charged
$\stauone$'s. The generation of a particle-antiparticle asymmetry
requires a departure from thermal equilibrium. Therefore, one might
think that a $\stauone$-$\stauone^*$ asymmetry can be produced at the
time of the stau freeze out if the (slepton number violating) process
$\stauone\stauone \rightarrow \tau\tau$ occurs at a different rate
than its conjugate counterpart.  Such a situation might indeed occur
if we allow for (CP-violating) complex values of the parameters
$\Atau$, $\mu$, and $M_{1,2}$ in the SUSY sector.  However, the staus
are still tightly coupled to Standard Model particles so that they
remain in kinetic equilibrium with the primordial plasma. Therefore,
any excess of $\stauone^*$ over $\stauone$ arising will be washed out
quickly by the inelastic scattering process
$\stauone^*\tau\leftrightarrow\stauone\overline{\tau}$.%
\footnote{We are grateful to M.~Pospelov for pointing out that
  process. Additional equilibrating processes are, e.g., $\stauone^*
  W^-\leftrightarrow \stauone W^+$ or
  $\stauone^*H^-\leftrightarrow\stauone H^+$, which are however
  Boltzmann-suppressed. Also note that a lepton asymmetry of the order
  of the baryon asymmetry is expected in baryogenesis scenarios based
  on leptogenesis.}
Indeed, it is well-known~\cite{Griest:1990kh} that processes of the
latter type occur at much larger rates than the rates for the mutual
annihilation of the decoupling particle species. The same argument
given in~\cite{Griest:1990kh} can be adopted to our case. At the time
of freeze out, the reaction rates of interest can be estimated as
\begin{align}
  \label{eq:p1}
  \stauone\stauone \rightarrow \tau \tau\, :&\quad n_{\stauone}
  n_{\stauone} \sigma_{\stauone\stauone\rightarrow\tau \tau} \sim
  T^3 \mstauone^3 
  e^{-2\mstauone/T} \sigma_{\stauone\stauone\rightarrow\tau \tau} \, ,\\
  \label{eq:p2}
  \stauone^*\tau \rightarrow \stauone \overline{\tau}\, :& \quad
  n_{\stauone^*} n_{\tau} \sigma_{\stauone^* \tau \rightarrow \stauone
    \overline{\tau}} \sim T^{9/2} \mstauone^{3/2} e^{-\mstauone/T}
  \sigma_{\stauone^* \tau \rightarrow \stauone \overline{\tau}}\, ,
\end{align}
since $\stauone^{(*)}$ is approximately Boltzmann distributed. For
simplicity, we have treated the tau lepton $\tau$ as a (still)
relativistic species. By taking the ratio of~(\ref{eq:p2}) with
respect to~(\ref{eq:p1}),
\begin{align}
  \left( T/\mstauone \right)^{3/2} e^{\mstauone/T} \sim 10^9\quad
  \mathrm{for}\quad \mstauone/T \simeq 25\ ,
\end{align}
we find that the equilibrating process is by far more dominant. Here,
we have used that $\sigma_{\stauone\stauone\rightarrow\tau \tau}$ and
$\sigma_{\stauone^* \tau \rightarrow \stauone \overline{\tau}}$ are
not too different. In fact, both processes proceed at tree level
exclusively via $\widetilde{\chi}_i^0$ exchange so that one cannot
decouple (\ref{eq:p2}) from (\ref{eq:p1}) by a simple adjustment of
the neutralino mass spectrum.

\section{Exceptionally small stau abundances within the CMSSM}
\label{sec:annih-chann}

We have shown above that the total stau annihilation cross section can
be significantly enhanced. The thermal freeze out of $\stauone$'s is
thereby delayed such that their abundance prior to decay, \Ystau, is
suppressed.  In the following we focus on the CMSSM to see whether the
effects discussed in Sects.~\ref{sec:enhanc-coupl-higgs}
and~\ref{sec:reson-annih} do appear also in models in which the
pattern of soft-SUSY breaking parameters fulfills certain boundary
conditions at a high scale.  Note that we compute \Ystau\ with
\texttt{micrOMEGAs} in this section since coannihilation processes are
not included in our relic density code. In addition, we employ
\texttt{SPheno~2.2.3}~\cite{Porod:2003um} for the computation of the
mass spectrum and the low energy constraints associated with
$B(b\rightarrow s \gamma)$ and the anomalous magnetic moment of the
muon $a_{\mu}$.  Let us now proceed by discussing two exemplary CMSSM
parameter scans.

\begin{figure}[t]
\begin{center}
\includegraphics[width=0.65\textwidth]{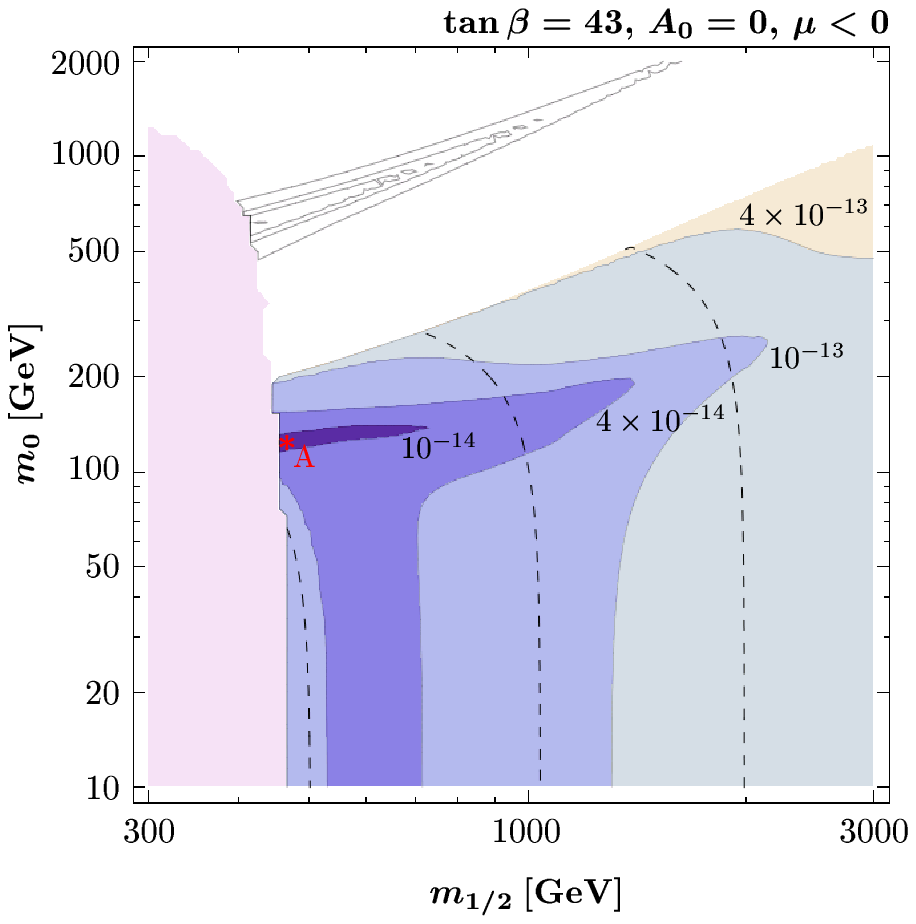}
\caption{\small Contours of $\Ystau$ (as labeled) in the
  (\monetwo,\,\mzero) plane for $\tanb=43$, $A_0=0$, and $\mu<0$,
  where darker shadings imply smaller $\Ystau$ values.  The dashed
  lines are contours of $\mstauone=100$, $300$, and $600\ \GeV$ (from
  left to right).  The light-shaded region at $\monetwo \lesssim
  450~\GeV$ is excluded by the LEP bound $\mh \le 114.4\
  \GeV$~\cite{Yao:2006px}.  In the white area either $m_{\neutralino}
  < \mstauone$ or correct electroweak symmetry breaking is not
  established (in the very upper left corner), where the thin contours
  indicate the Higgs funnel in the $\neutralino$ LOSP region.
  Table~\ref{tab:points} provides detailed information for the SUSY
  model represented by the point ``A'' that is indicated by the star.}
\label{Fig:cmssm1}
\end{center}
\end{figure}

Figure~\ref{Fig:cmssm1} shows contours of constant $\Ystau$ in the
(\monetwo,\,\mzero) plane for $\tanb =43$, $A_0 = 0$, and a negative
sign of the $\mu$ parameter.
%
%
%
The contour lines represent the values $\Ystau=10^{-14}$, $4\times
10^{-14}$, $10^{-13}$, and $4\times 10^{-13}$, where darker shadings
imply smaller values of $\Ystau$.
The dashed lines are contours of $\mstauone = 100$, $300$ and
$600~\GeV$ (from left to right). The light-shaded region at $\monetwo
\lesssim 450~\GeV$ is excluded by the mass bound $\mh \ge 114.4\ \GeV$
from Higgs searches at LEP~\cite{Yao:2006px}. The white area indicates
the region in which either correct electroweak symmetry breaking is
not established (in the very upper left corner) or in which
$m_{\neutralino}<\mstauone$. Since $\mu<0$, the plane is actually in
tension because of (negative) SUSY contributions
$a^{\mathrm{SUSY}}_\mu$ to the anomalous magnetic moment of the muon,
$a_\mu\equiv(g-2)_\mu/2$.

\begin{figure}[t]
\begin{center}
\includegraphics[width=0.65\textwidth]{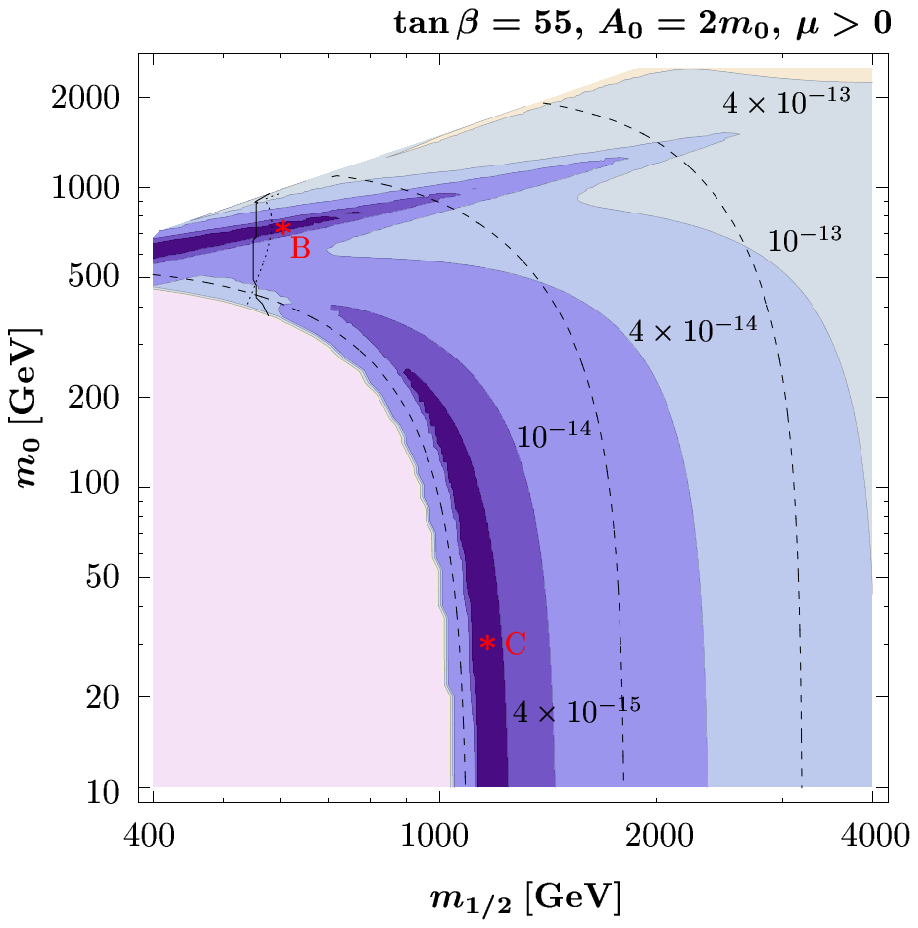}
\caption{\small Contours of $\Ystau$ (as labeled) in the
  (\monetwo,\,\mzero) plane for $\tanb =55$, $A_0 = 2m_0$, and
  $\mu>0$, where darker shadings imply smaller $\Ystau$ values. The
  dashed lines are contours of $\mstauone = 100,\, 300,$ and $600\
  \GeV$ (from left to right).  The large light-shaded region in the
  lower left corner is excluded by bounds from direct Higgs and SUSY
  searches (or by the appearance of a tachyonic spectrum). In the
  region to the left of the vertical solid and dotted lines, $\mh \le
  114.4\ \GeV$~\cite{Yao:2006px} and $B(b \rightarrow s \gamma)\ge
  4.84\times 10^{-4}$~\cite{Mahmoudi:2007gd}, respectively.  In the
  white area, $m_{\neutralino} < \mstauone$. Table~\ref{tab:points}
  provides detailed information for the SUSY models represented by the
  stars ``B'' and ``C'' (as labeled).}
\label{Fig:cmssm2}
\end{center}
\end{figure}

Figure~\ref{Fig:cmssm2} presents a scan over the (\monetwo,\,\mzero)
plane for $\tanb =55$, $A_0 = 2m_0$, and $\mu>0$
%
%
with contours of $\Ystau=4\times 10^{-15}$, $10^{-14}$, $4\times
10^{-14}$, $10^{-13}$, and $4\times 10^{-13}$ (darker shadings
indicate smaller $\Ystau$ values) and $\mstauone=100$, $300$, and
$600~\GeV$ (dashed lines, from left to right). The large light-shaded
region in the lower left corner is excluded by the robust bound
$\mstauone\ge 82\ \GeV$~\cite{Yao:2006px} from collider searches of
charged sleptons (or by the appearance of a tachyonic spectrum).  The
LEP Higgs bound $\mh \le 114.4\ \GeV$~\cite{Yao:2006px} is situated
within this region in close vicinity to its boundary for
$\mzero\lesssim 400~\GeV$ and is indicated by the solid line for
$\mzero\gtrsim 400~\GeV$. In the region to the left of the dotted
line, $B(b\rightarrow s\gamma)\ge 4.84\times
10^{-4}$~\cite{Mahmoudi:2007gd}, which is in tension with the bounds
from inclusive $b\,\rightarrow\,s\gamma$ decays.

Let us now discuss some generic features of the stau yield within the
CMSSM on the basis of Figs.~\ref{Fig:cmssm1} and~\ref{Fig:cmssm2}.  We
note beforehand that our more general statements on the $\stauone$
LOSP region in the CMSSM are corroborated by a parameter scan over the
following range%
\footnote{Here, we disregard CMSSM parameter points in which \texttt{SPheno}
  flags an error in the spectrum calculation.}
\begin{alignat}{2}
  \label{eq:para-range}
  &\monetwo = (0.1 - 6)\ \TeV,\qquad &
  \tanb &= 2 - 60, \nonumber\\
   &-4 \mzero < A_{0} < 4 \mzero  ,
   &\mathrm{sgn\ \!}\mu &= \pm 1 .
\end{alignat}
In both figures an almost horizontal, narrow band of low $\Ystau$
appears in which $2\mstauone\simeq\mH$ holds so that stau annihilation
proceeds via resonant production of the heavy CP-even Higgs boson
\Hhiggs.
We have marked points the centers of the respective regions with ``A''
and ``B'' for which we provide detailed information in
Table~\ref{tab:points}.  Given a present uncertainty of $\sim 3\ \GeV$
in the determination of $m_{\hhiggs}$~\cite{Degrassi:2002fi}, we note
that the LEP Higgs bound has to be treated with some care. For
example, a (vertical) $m_{\hhiggs}=112\ \GeV$ contour would be
situated at $\monetwo\simeq 400\ \GeV$ in the resonance region of
Fig.~\ref{Fig:cmssm2}. Accordingly, one could consider the entire
resonance region shown to be compatible with direct Higgs searches.
However, due to the large value of $\tanb=55$, the bound on
$b\,\rightarrow\,s\gamma$ is very strong so that a large part of the
resonance region remains excluded by this constraint. In this regard,
it is interesting to see (Fig.~\ref{Fig:cmssm1}) that
$2\mstauone\simeq\mH$ also appears in the $\stauone$ LOSP region for
lower values of $\tanb$.  In the center of both resonance regions, the
yield becomes as low as $\Ystau= 4.2\times 10^{-15}$ (point A) and
$\Ystau=2.5\times 10^{-15}$ (point B). Despite the heavier mass of the
lighter stau (see Table~\ref{tab:points}), the suppression of $\Ystau$
is still more pronounced in Fig.~\ref{Fig:cmssm2} than in
Fig.~\ref{Fig:cmssm1}.  This is because the bottom Yukawa coupling
becomes larger with increasing $\tanb$, as discussed already in
Sect.~\ref{sec:reson-annih}. In fact, annihilation into
$b\overline{b}$ final states is in both cases by far the dominant
process with relative importances of $76\%$ (point A) and $87\%$
(point B). The extension of both resonance regions is due to the total
width of $\Hhiggs$ of respectively $\Gamma_{\Hhiggs}\simeq 9.6~\GeV$
(point A) and $\Gamma_{\Hhiggs}\simeq 22~\GeV$ (point B); note the
logarithmic scales in Figs~\ref{Fig:cmssm1} and~\ref{Fig:cmssm2}.  We
note in passing that the appearance of the $\Hhiggs$ resonance does
not imply the absence of the neutralino funnel region which is
indicated by the (unshaded) contour lines in the white area of
Fig.~\ref{Fig:cmssm1}

Of course, the question arises whether the appearance of the resonance
region is encountered more generically within the framework of the
CMSSM.
In principle, it is not easy to provide a simple quantitative
connection between $\mstauone$ and $\mH$ for arbitrary values of the
CMSSM parameters. However, without emphasis on an overall
applicability, a qualitative picture can be drawn. Let us start with
the mass of the CP-odd Higgs boson $\mA$ which can be written
as~\cite{Drees:1991mx,Drees:1995hj}
\begin{align}
  \label{eq:mA}
  \mA^2 \sim 1/\sin^{2}{\beta}\, (\mzero^2 + 0.52 \monetwo^2 + \mu^2  - \dots) .
\end{align}
Here, the ellipsis stand for contributions from the bottom and tau
Yukawa couplings. For $\tanb\gtrsim 20$, $\mA^2 \sim
\mzero^2+2.5\monetwo^2-\dots$, and the corrections from the bottom and
tau Yukawa couplings become important so that $\mA$ is driven towards
lower values;%
\footnote{The latter relation ignores contributions from $A$-terms
  which can be important but complicate the envisaged illustrative
  picture; for the derivation, we have used
  $m_{\mathrm{t}}(m_{\mathrm{t}})=163\ \GeV$ in Eq.~(2.25a) of
  Ref.~\cite{Drees:1995hj}.}
note that $\sin^{2}{\beta}\simeq1$ for $\tanb\gtrsim 20$. Indeed, this
property can be used to constrain $\tanb$ from above by confronting
$\mA$ with the lower bound from LEP, $\mA>93.4\
\GeV$~\cite{Yao:2006px}. On the other hand, for large $\monetwo$, one
also enters the decoupling limit of the MSSM so that $\mA$ and $\mH$
will be nearly degenerate in mass; cf.~(\ref{eq:DL-higgs-masses}).
This can be also seen from the exemplary points presented in
Table~\ref{tab:points}. Therefore, also $\mH$ will be driven towards
lower values for growing $\tanb$.  Now, left-right mixing of the
lighter stau for not too large values of $\tanb$ is small within the
CMSSM, $\stauone\simeq\stauR$, so that approximately $\mstauone^2 \sim
\mzero^2 + 0.15\monetwo^2$~\cite{Drees:1995hj}.  Therefore,
$2\mstauone <\mH$ is the relation that holds usually in the region in
which $\stauone$ is the lightest Standard Model superpartner.
However, for large $\tanb$, the contributions from the bottom Yukawa
coupling in~(\ref{eq:mA}) can become strong enough (growing with
$\mzero$~\cite{Drees:1995hj}) to overcome any additional decrease of
$\mstauone$ due to left-right mixing so that the resonance condition
$2\mstauone\simeq\mH$ can indeed be met. Nevertheless, from scanning
over the CMSSM parameter range (\ref{eq:para-range}) it seems to us
that the resonance condition $2\mstauone\simeq\mH$ is not easily
realized in the part of the $\stauone$ LOSP region in which
$\stauone$-$\neutralino$ coannihilations are negligible.  Conversely,
it is clear that relaxing the universality conditions for the soft-SUSY
breaking masses at $\mgut$ will make it easier to find parameter
regions in which the resonance condition $2\mstauone\simeq\mH$ is
satisfied. Of particular interest in this respect is the model with
non-universal Higgs masses (NUHM) with $m_{H_1}\neq m_{H_2} \neq m_0$
at $\mgut$. There, one can adjust the input parameters in order to
realize resonant stau annihilation.  Indeed, this model is
qualitatively the same as the class of pMSSM scenarios considered in
the previous sections, where $m_{H_1}$ and $m_{H_2}$ are traded (at
the low-scale) against $\mA$ and $\mu$ by using the electroweak
symmetry breaking conditions.

Low $\Ystau$ values are also realized in the narrow vertical region
around $\monetwo\sim 1.1\ \TeV$ in Fig.~\ref{Fig:cmssm2}. At the
representative point ``C'' of that region, $\Ystau = 2.2\times
10^{-15}$ and the main stau annihilation channels are the ones into
$\hhiggs\hhiggs$ ($90\%$) and $WW$ ($6\%$); see
Table~\ref{tab:points}.  For larger values of $\monetwo$, \Ystau\
exhibits its well known behavior and grows with $\mstauone$. To the
left of the $\Ystau=4\times 10^{-15}$ contour, the yield increases
quickly since the annihilation into $\hhiggs\hhiggs$ becomes
kinematically forbidden. Indeed, regions of low $\Ystau$ which are due
to the aforementioned annihilation channels are a commonplace
appearance in the CMSSM parameter space.  They are found slightly
above the lowest feasible values of $\monetwo$, i.e., close to the
boundary of the region which is excluded by direct Higgs and SUSY
searches and where $\mstauone>\mh$ still holds.  This is because
$\stauone$ is light in that region since the SUSY particle spectrum
scales with $\monetwo$ (typically, $m_0\ll\monetwo$ for $\stauone$
LOSP).  Moreover, we find that the LEP Higgs bound drops hardly below
$\monetwo\simeq 450\ \GeV$ for $\tanb\gtrsim 40$ and $\mzero\lesssim
100\ \GeV$.%
\footnote{The position of the LEP Higgs bound (which appears as a near
  to vertical line for low \mzero) is very sensitive to the value of
  $\mtop$. Lowering $\mtop$ shifts the bound towards larger values of
  \monetwo.}
Due to a strong correlation between the gaugino mass parameter
$\monetwo$ and the size of the $\mu$ parameter, $\mu^2\sim (1- 3)\,
\monetwo^2$~\cite{Carena:1994bv}, the value of $\mu$ in the
experimentally allowed region is large.  Recall from
Sect.~\ref{sec:enhanc-coupl-higgs} that the $\stauone\stauone\hhiggs$
coupling is $\sim\sin{2\thetastau}\Xtau$ ($\mA\gg \mZ$) so that
$|\Xtau|=|\Atau-\mu\tanb|$ will become sizeable by increasing $\tanb$.
This leads then to efficient stau annihilation into $\hhiggs\hhiggs$
final states.
Indeed, in those CMSSM regions, also $|\sin{2\thetastau}|$ is
maximized so that $\Ystau$ already starts to drop below the estimate
(\ref{eq:yield-approx}) for $\tanb\gtrsim 40$.
Note, however, that the left-right mixing of $\stauone$ within the
CMSSM is somewhat constrained. Neglecting $\tau$-Yukawa contributions,
the RG-evolution induced splitting reads $\mstauL^2-\mstauR^2\sim
0.37\monetwo^2$~\cite{Drees:1995hj} and indeed $\stauone$ remains
mainly right-handed: By scanning over the parameter space, we
typically find $65^\circ\lesssim\thetastau\lesssim 115^\circ$ and thus
$|\sin{2\thetastau}|\lesssim 0.75$ in the $\stauone$ LOSP region in
which $\mstauone>\mh$ and $\mh>114.4\ \GeV$ holds.

%
\begin{table}[tb] 
  \caption{\small Exemplary CMSSM points A, B, and C shown in Figs.~\ref{Fig:cmssm1} and \ref{Fig:cmssm2}. In addition to the quantities explained in the main text, values of the gluino mass $m_{\widetilde{g}}$ and of the mass of the lighter stop $m_{\widetilde{t}_1}$ are given together with the relative importance of the dominant stau annihilation channels, $x_{\mathrm{f}}=\mstauone/\Tf$, and the decoupling yield $\Ystau$. For each point, we list gravitino dark matter scenarios with $\mgr = 100\ (50)\ \GeV$ and associated values of the stau lifetime $\taustau$, the non-thermally produced gravitino density $\Omega_{\gravitino}^{\NTP}h^2$, and the maximum reheating temperature $\TR^{\mathrm{max}}$.}
\label{tab:points} 
\begin{center}
\begin{tabular}{l@{}l@{\hspace*{0.5cm}}ccc}
\toprule\\[-0.4cm]
Point & & A & B & C  \\[0.1cm]
\midrule\\[-0.50cm]
\monetwo & $[\GeV]$           &   456     &   600   &  1138   \\[0.1cm] 
\mzero &$[\GeV]$              &   124     &   748   &  30     \\[0.1cm]
\tanb  &                      &   43      &   55    &  55     \\[0.10cm]
\midrule\\[-0.40cm]
\mstauone &$[\GeV]$           &   130     &   197   &  127    \\[0.1cm]
\mstautwo &$[\GeV]$           &   352     &   673   &  739    \\[0.1cm]
\thetastau &                  &   114     &   80    &  75     \\[0.10cm]
\midrule\\[-0.40cm]                               
$m_{\hhiggs}$ & $[\GeV]$        &   114.6     &   115   &  117.9   \\[0.1cm] 
$m_{\Hhiggs\! , \Ahiggs}$ &  $[\GeV]$ &   265     &   390   &  799     \\[0.1cm]
$\Gamma_{\Hhiggs}$& $[\GeV]$    &   9.6      &   22    &  41     \\[0.1cm]
$\mu$& $[\GeV]$               &   -565     &   666   &  1262    \\[0.1cm]
\Atau & $[\GeV]$              &   -63     &   473   &  -164    \\[0.10cm]
\midrule\\[-0.40cm]
$m_{\widetilde{g}}$& $[\GeV]$     &   1052       &   1375  &  2446   \\[0.1cm] 
$m_{\widetilde{t}_1}$ & $[\GeV]$   &   740       &   1091  &  1757   \\[0.10cm] 
\midrule\\[-0.40cm]                               
$b\overline{b}$   &   $[\%]$   &   76    &   87   &   $<1$   \\[0.1cm] 
$\hhiggs\hhiggs$  &   $[\%]$  &   10    &   $<1$   &   90   \\[0.1cm] 
$\tau\overline{\tau}$ & $[\%]$  &   9    &    11   &   $<1$   \\[0.1cm] 
$WW$   &   $[\%]$             &   2    &   $<1$   &   6   \\[0.1cm] 
$x_{\mathrm{f}}$ &            &   30      &   30    &  32   \\[0.1cm] 
\Ystau & $[10^{-15}]$         &   4.2    &   2.5   &   2.2   \\[0.10cm] 
\midrule\\[-0.40cm]                               
\mgr & $[\GeV]$               &   100    &   100    &   100    \\[0.1cm]
     &                        &   (50)    &   (50)    &  (50)    \\[0.2cm]
\taustau  & $[\mathrm{s}]$ &  $5.7\times 10^9$  &  $6.5\times 10^7$  &  $8.5\times 10^9$  \\[0.1cm]
          &                & $(7.5\times 10^7)$ & $(6.4\times 10^6)$ & $(8.7\times 10^7)$ \\[0.2cm]
$\Omega^{\mathrm{NTP}}_{\widetilde{G}}h^2$ & $[10^{-4}]$ & $1.2$ &  $0.7$ & $0.64$ \\[0.1cm]
                                        &&   ($0.58$)  & ($0.35$) &  ($0.32$) \\[0.2cm]
$\TR^{\mathrm{max}}$ & $[\GeV]$ & $1.9\times 10^9$ &  $1.1\times 10^9$          &  $3.1\times 10^8$ \\[0.1cm]
                     &          & ($9.5\times 10^8$)       & ($5.5\times 10^8$) & ($1.5\times 10^8$)    \\[0.2cm]
\bottomrule

\end{tabular}
\end{center}
\end{table}
%

\section{Prospects for collider phenomenology}
\label{sec:collider}

If a SUSY model with a long-lived $\stauone$ of $\mstauone<0.7~\TeV$
is realized in nature, the $\stauone$ discovery potential will be
promising at the LHC with a luminosity of
100~fb$^{-1}$~\cite{Feng:2004mt}.
For $\mstauone<0.25~\TeV$ (0.5~TeV), $\stauone$'s can also be examined
in precision studies at the ILC with a c.m.\ range up to
$\sqrt{s}=0.5~\TeV$ (1~\TeV).
Once long-lived $\stauone$'s are produced, one should be able to
distinguish them from muons by considering the associated highly
ionizing tracks and with time-of-flight measurements.
One should then also be able to infer $\mstauone$ from measurements of
the $\stauone$ velocity and its momentum~\cite{Ambrosanio:2000ik} and
complementary from (threshold) studies of the process
$\positron\electron\to\stauone\stauone^*$ at the ILC.

Remarkably, both mechanisms leading to exceptionally small $\Ystau$
values come with testable predictions: certain ranges of the
stau-mixing angle $\thetastau$ together with large values of $\tanb$,
$|\mu|$, and/or $|\Atau|$ and, in the case of resonant stau
annihilation, also $\mH\simeq 2\mstauone$.
In particular, the large stau-Higgs couplings lead to an enhanced
production of light Higgs bosons in association with staus via
$\positron\electron\to\stauone\stauone^*\hhiggs$ and
$\gamma\gamma\to\stauone\stauone^*\hhiggs$.  The associated cross
sections can then be relatively large at the ILC with a sufficiently
high c.m.\ energy~\cite{Datta:2001sh}.  In addition, the above
reactions with $\Hhiggs$ instead of $\hhiggs$ in the final state can
have also relatively large cross sections if $\Hhiggs$ and $\stauone$
are sufficiently light.  These reactions will then allow for an
experimental determination of the stau-Higgs couplings and clarify
whether its values are compatible with an extremely small value of
$\Ystau$~\cite{Datta:2001sh}.  Moreover, a measurement of $\mH$
pointing to $\mH\simeq 2\mstauone$ could be an experimental hint for
resonant stau annihilation in the early Universe.

Indeed, the scenarios considered could allow for a determination of
both $\mh$ and $\mH$ already at the LHC.  Because of the large values
of $\tanb$, the dominant production mechanism for $\hhiggs/\Hhiggs$
will be the associated production of the neutral Higgs bosons with
bottom quark pairs, $\proton\proton\to
\bquark\antibquark\hhiggs/\Hhiggs$; see,
e.g.,~\cite{Dittmaier:2003ej,Dawson:2003kb,Harlander:2003ai,Dittmaier:2006cz}
and references therein.  In fact, associated
$\bquark\antibquark\hhiggs/\Hhiggs$ production with
$\hhiggs/\Hhiggs\to\mu^+\mu^-$ is considered as one of the most
promising processes for measurements of $\mH$ at the LHC despite the
relatively small $\hhiggs/\Hhiggs\to\mu^+\mu^-$ branching
ratio~\cite{Ball:2007zza}.  In SUSY scenarios with a sufficiently light
long-lived $\stauone$ LOSP, these processes will be complemented by
associated $\bquark\antibquark\hhiggs/\Hhiggs$ production with
$\hhiggs/\Hhiggs\to\stauone\stauone^*$, where measurements of the
invariant mass of the $\stauone\stauone^*$ pair could potentially
provide a unique way to infer $\mh$ and $\mH$ at the LHC. In fact,
$\hhiggs/\Hhiggs\to\stauone\stauone^*$ will occur most prominently
exactly in the regions associated with the exceptional $\Ystau$ values
due to the enhanced stau--Higgs couplings. Having outlined these
proposals, we leave a dedicated study for future work.

Table~\ref{tab:points} illustrates that the kinematical reach of both
the LHC and the ILC could be sufficiently large to allow for the
studies mentioned above.  In none of the given points does $\mstauone$
exceed $200~\GeV$ so that $\stauone\stauone^*$ pair production would
already be possible at the ILC with $\sqrt{s}\leq 0.5~\TeV$.  There,
one could also produce $\stauone\stauone^*\hhiggs$ final states in
scenarios A and C.  Even the condition $\mH\simeq 2\mstauone$ could be
probed in both scenarios A and B that allow for resonant stau
annihilation.

\section{Implications for gravitino dark matter scenarios}
\label{sec:gravitino}

As already stressed in the Introduction, $\Ystau$ is subject to
stringent cosmological constraints. Indeed, to decide on the
cosmological viability of a SUSY model, one has to confront the
associated $\Ystau$ values with those constraints. In particular, for
gravitino LSP scenarios with unbroken R-parity, restrictive
cosmological constraints and implications thereof have been
derived~\cite{Cyburt:2006uv,Steffen:2006wx,Pradler:2006hh,Pradler:2007is,Kersten:2007ab,Pradler:2007ar,Kawasaki:2008qe,Steffen:2008bt,Pospelov:2008ta}
often under the assumption that $\Ystau$ can be described
by~(\ref{eq:yield-approx}). However, while (\ref{eq:yield-approx}) is
quite reliable for
$\stauone\simeq\stauR$~\cite{Asaka:2000zh,Fujii:2003nr,Pradler:2006hh,Berger:2008ti},
we have shown in the previous sections that $\Ystau$ (for a given
$\mstauone$) can be about than two orders of magnitude smaller
than~(\ref{eq:yield-approx}). Thus, in gravitino dark matter scenarios
with such exceptionally small $\Ystau$ values, our understanding of
the cosmological constraints and the associated implications could
change significantly.

To demonstrate this point, let us indicate for which $\Ystau$ values
the existing cosmological constraints (in their present status) are
respected:
\begin{itemize}
\item For $\Ystau<10^{-14}$, the upper limit on $\Ystau$ imposed by
  the non-thermal production of gravitinos in $\stauone$ decays,
  $\Omega_{\gravitino}^{\NTP}\leq f\,\Omega_{\mathrm{dm}}$---given
  explicitly in~(22) of Ref.~\cite{Steffen:2006hw}---is respected for
  $\mgr\lesssim 500~\GeV$ even if only a small fraction $f=0.01$ of
  dark matter is assumed to originate from $\stauone$ decays; cf.\
  Fig.~13 of Ref.~\cite{Steffen:2006hw}.
  This applies equally to other scenarios with an extremely weakly
  interacting LSP---such as the axino
  LSP~\cite{Covi:1999ty,Brandenburg:2005he}---originating from
  $\stauone$ decays.
\item For $\Ystau\lesssim 10^{-13}$, the BBN constraints associated
  with effects of hadronic energy release on the primordial D
  abundance can be respected for $\stauone\simeq\stauR$ and
  $\mstauone$ up to $10~\TeV$ independent of the $\stauone$ lifetime;
  cf.~Fig.~11 of Ref.~\cite{Steffen:2006hw}. For a sizable admixture
  of $\stauL$ in $\stauone$, this $\Ystau$ constraint can become more
  restrictive in particular with the enhanced stau--Higgs couplings
  allowing for exceptionally small $\Ystau$ values. Nevertheless,
  these exceptional values are typically associated with $\mstauone<
  300~\GeV$ where the $\Ystau$ limit is significantly more relaxed:
  $\Ystau\lesssim 10^{-11}$ for $\stauone\simeq\stauR$.
  A tightening to $\Ystau\lesssim 10^{-13}$ ($10^{-15}$) will then
  require an increase of $\epsilon_{\mathrm{had}}$ by a factor of
  $10^2$ ($10^4$).
  On the other hand, sufficiently degenerate $\mgr$ and $\mstauone$
  will always be associated with small values of
  $\epsilon_{\mathrm{had}}$ and thereby with relaxed $\Ystau$ limits
  from energy release, even in the case of strongly enhanced
  stau--Higgs couplings.
\item For $\Ystau\lesssim 10^{-14}~(10^{-15})$, the BBN constraints
  associated with effects of electromagnetic energy release on the
  primordial D ($^3$He) abundance can be respected independent of the
  $\stauone$ lifetime; cf.~upper panels of Fig.~12
  ($100~\GeV\leq\mstauone\leq 10~\TeV$) of Ref.~\cite{Steffen:2006hw}
  and Figs.~14 ($\mstauone=100~\GeV$) and~15 ($\mstauone=300~\GeV$) of
  Ref.~\cite{Kawasaki:2008qe}.
\item For $\Ystau\lesssim 2\times 10^{-15}~(2\times 10^{-16}-2\times
  10^{-15})$, the BBN constraints associated with bound state effects
  allowing for CBBN of $^9$Be ($^6$Li) can be respected even for
  $\taustau\gtrsim 10^5\,\seconds$; cf.\ Fig.~5 in
  Ref.~\cite{Pospelov:2008ta}. The uncertainty on the $\Ystau$ limit
  associated with CBBN of $^6$Li reflects the difficulties in
  quantifying an upper limit on the primordial $^6$Li abundance; for
  more details on this issue, see~\cite{Pospelov:2008ta} and
  references therein.
  Note that the given limits correspond to upper limits on the
  primordial fractions of $^9$Be/H and $^6$Li/H of $2.1\times
  10^{-13}$ and $10^{-11}-10^{-10}$, respectively.
\end{itemize}
Thus, the SUSY models which come with thermal relic stau abundances of
$\Ystau \lesssim 2\times 10^{-15}$ can respect each of these
cosmological constraints independent of the stau lifetime if a
primordial $^6$Li/H abundance of about $10^{-10}$ is viable. In
particular, the limit of $\tau_{\stauone}\lesssim 5\times
10^3\,\seconds$ and its implications---discussed in the
Introduction---are then no longer valid even for a standard
cosmological history with primordial temperatures of $T>\Tf$. Thereby,
the regions with $\Ystau \lesssim 2\times 10^{-15}$ are associated
with particularly attractive gravitino dark matter scenarios:
\begin{itemize}
\item The gravitino mass can be within the range
  $0.1\lesssim\mgr<\mstauone$ for which its kinematical determination
  could be
  viable~\cite{Buchmuller:2004rq,Martyn:2006as,Hamaguchi:2006vu}.
  Together with measurements of $\mstauone$ and $\taustau$, a
  kinematically determined $\mgr$ would allow one to measure the
  Planck scale $\MPl$ at
  colliders~\cite{Buchmuller:2004rq,Martyn:2006as,Hamaguchi:2006vu}.
  Indeed, an agreement of the $\MPl$ value determined in collider
  experiments with the one inferred from Newton's constant $G_{\rm N}$
  would support the existence of supergravity in
  nature~\cite{Buchmuller:2004rq}.
\item For $\mgr$ sufficiently close to $\mstauone$, the spin 3/2
  character of the gravitino becomes relevant so that it could be
  probed in principle by analyzing the decays
  $\stauone\to\gravitino\tau\gamma$~\cite{Buchmuller:2004rq}.
\item With $\Ystau \lesssim 2\times 10^{-15}$,
  $\Omega_{\gravitino}^{\NTP}$ is negligible so that
  basically all of $\Omega_{\mathrm{dm}}$ can be provided by
  gravitinos from other sources such as thermal
  production~\cite{Bolz:2000fu,Pradler:2006qh,Rychkov:2007uq}.
  Indeed, if also gravitino production in decays of scalar fields such
  as the inflaton~\cite{Asaka:2006bv,Endo:2007sz} is negligible,
  reheating temperatures of $\TR\lesssim 10^9\,\GeV$ could become
  viable for $\mgr\sim 100~\GeV$ and not too heavy gaugino
  masses~\cite{Pradler:2006hh,Rychkov:2007uq}.
  This would mean that thermally produced gravitinos could provide the
  right amount of dark matter and that thermal
  leptogenesis~\cite{Fukugita:1986hr} with hierarchical heavy Majorana
  neutrinos---which typically requires $\TR\gtrsim
  10^9\,\GeV$~\cite{Davidson:2002qv,Buchmuller:2004nz,Blanchet:2006be,Antusch:2006gy}---would
  be a viable explanation of the cosmic baryon asymmetry, i.e., there
  would be no gravitino problem.
\item With a kinematically determined $\mgr$, one would be able to
  probe the reheating temperature $\TR$ at colliders and thereby the
  viability of thermal leptogenesis~\cite{Pradler:2006qh}.
\item For $\taustau\gtrsim 10^4\,\seconds$, the small $\Ystau$ values
  could still allow for the primordial catalysis of $^6$Li and $^9$Be
  in agreement with existing astrophysical
  observations~\cite{Pospelov:2006sc,Bird:2007ge,Pospelov:2007js,Pospelov:2008ta}.
\end{itemize}

Table~\ref{tab:points} illustrates that gravitino dark matter
scenarios of the type discussed above can even be accommodated within
the CMSSM.%
\footnote{Note that this finding points to a caveat of our earlier
  study~\cite{Pradler:2007is} relying on $\taustau\lesssim 5\times
  10^3\,\seconds$ derived from the assumption that $\Ystau$ is
  described by~(\ref{eq:yield-approx}). Our previous sections however
  show that islands exist in which $\Ystau$ can be significantly
  below~(\ref{eq:yield-approx}) even within the CMSSM and for a
  standard cosmological history.}
For gravitino masses of $50~\GeV$ and $100~\GeV$, we list the
associated values of $\taustau$, of $\Omega_{\gravitino}^{\NTP}h^2$,
and of the maximum reheating temperature $\TR^{\mathrm{max}}$ under
the assumption that other gravitino sources can be neglected.  The
expression for $\taustau$ can be found, e.g., in (8) of
Ref.~\cite{Pradler:2007is} whereas (12) of the same reference can be
used to infer the $\TR^{\mathrm{max}}$ values imposed by
$\Omega_{\gravitino}^{\TP}h^2\leq 0.126$.  At each CMSSM point and for
both $\mgr$ values, $\stauone$ is very long lived, $\taustau>
10^6\,\seconds$, and gravitino production from $\stauone$ decays is
negligible, $\Omega_{\gravitino}^{\NTP}h^2\lesssim 10^{-4}$.  In all
cases, the gravitino mass $\mgr=100~\GeV$ is sufficiently close to
$\mstauone$ so that the spin 3/2 character of the gravitino can in
principle be probed~\cite{Buchmuller:2004rq}.  A reheating temperature
of $\TR\gtrsim 10^9\,\GeV$ is viable only for the points A and B with
$\monetwo$ significantly below $1~\TeV$, i.e., at the points at which
resonant stau annihilation leads to the reduction of $\Ystau$.
Because of $\taustau> 10^6\,\seconds$, the $\Ystau$ limit from CBBN of
$^9$Be is at $Y_{\mathrm{Be}}^{\mathrm{max}}\simeq 2\times 10^{-15}$
for each point as can be inferred from Fig.~5 of
Ref.~\cite{Pospelov:2008ta}.  This bound disfavors point A while the
points B and C are associated with $\Ystau$ values very close to this
limit and thereby with $^9\mathrm{Be/H}$ ($^6\mathrm{Li/H}$) values of
about $2.1\times 10^{-13}$ ($10^{-10}$)~\cite{Pospelov:2008ta}.

\section{Conclusions}
\label{sec:conclusions}

Supersymmetric models with a long-lived stau $\stauone$ being the
lightest Standard Model superpartner are well-motivated and very
attractive in light of potentially striking signatures at colliders.
For a standard thermal history with primordial temperatures
$T>\mstauone/20>\Tf$---which is the working hypothesis is this
work---the long-lived $\stauone$ becomes an electrically charged
thermal relic whose abundance can be restricted by cosmological
constraints.

We have carried out a thorough study of primordial stau annihilation
and the associated thermal freeze out.
Taking into account the complete set of stau annihilation channels
within the MSSM with real parameters for cases with negligible
sparticle coannihilation,
the resulting thermal relic $\stauone$ yield $\Ystau$ has been
examined systematically.
While related earlier studies focussed mainly on the
$\stauone\simeq\stauR$
case~\cite{Asaka:2000zh,Fujii:2003nr,Pradler:2006hh,Berger:2008ti}, we
have investigated cases in which $\stauone$ contains a significant
admixture of $\stauL$ including the maximal mixing case and
$\stauone\simeq\stauL$.

We find that the variation of the stau mixing angle $\thetastau$ does
affect the relative importance of the different annihilation channels
significantly but not necessarily the resulting $\Ystau$ value for
relatively small values of $\tanb$.
By increasing $\tanb$, however, we encounter a dramatic change of this
picture for large absolute values of the Higgs-higgsino mass parameter
$\mu$ and/or of the trilinear coupling $\Atau$, which are the
dimensionful SUSY parameters that govern simultaneously stau
left-right mixing and the stau--Higgs couplings: Stau annihilation
into $\hhiggs\hhiggs$, $\hhiggs\Hhiggs$, and $\Hhiggs\Hhiggs$ can
become very efficient (if kinematically allowed) so that $\Ystau$ can
decrease to values well below $10^{-15}$. 
The scalar nature of $\stauone$ allows those parameters to enter
directly into the annihilation cross sections. This mechanism has no
analogue in calculations of the thermal relic density of the lightest
neutralino $\neutralino$.

The stau--Higgs couplings are crucial also for the second $\Ystau$
reduction mechanism identified in this work:
Even for moderate values of $\tanb$, we find that staus can annihilate
very efficiently into a $\bquark\antibquark$ pair via $s$-channel
exchange of the heavy CP-even Higgs boson $\Hhiggs$ provided the MSSM
spectrum exhibits the resonance condition $2\mstauone\simeq\mH$.
We have shown explicitly that the associated $\Ystau$ values can be
below $10^{-15}$ as well.
This mechanism is similar to the one that leads to the reduction of
the $\neutralino$ density in the Higgs funnel region in which
neutralino annihilation proceeds at the resonance of the CP-odd Higgs
boson $\Ahiggs$.

We have worked with an effective low energy version of the MSSM to
investigate the $\thetastau$-dependence of $\Ystau$ and the two
$\Ystau$-reduction mechanisms in a controlled way. In addition, we
have shown that the considered effects can be accommodated also with
restrictive assumptions on the soft-SUSY breaking sector at a high
scale. Within the CMSSM, we encounter both mechanisms each of which
leading to $\Ystau\simeq 2\times 10^{-15}$ in two distinct regions of
a single $(\monetwo,\,\mzero)$ plane.

We have discussed possibilities to probe the viability of the
presented $\Ystau$-reduction mechanisms at colliders.  While a $\mH$
measurement pointing to $\mH\simeq 2\mstauone$ would support resonant
primordial stau annihilation, studies of Higgs boson production in
association with staus, $\positron\electron\,(\gamma\gamma)
\to\stauone\stauone^*\hhiggs,\stauone\stauone^*\Hhiggs$ could allow
for an experimental determination of the relevant stau--Higgs
couplings, for example, at the ILC.
Moreover, we have outlined that associated
$\bquark\antibquark\hhiggs/\Hhiggs$ production with
$\hhiggs/\Hhiggs\to\stauone\stauone^*$ has the potential to allow for
a determination of both $\mh$ and $\mH$ at the LHC if a SUSY scenario
with large $\tanb$ and large stau--Higgs couplings is realized.

With the obtained small $\Ystau$ values, even the restrictive
constraints associated with CBBN could be respected so that attractive
gravitino dark matter scenarios could be revived to be cosmologically
viable even for a standard cosmological history. Within this class of
models, collider evidence for supergravity, for the gravitino being
the LSP, and for high values of the reheating temperatures of up to
$10^9\,\GeV$ is conceivable, which could thereby accommodate
simultaneously the explanation of the cosmic baryon asymmetry provided
by thermal leptogenesis and the hypothesis of thermally produced
gravitinos being the dark matter in our Universe.

\bigskip

{\bf Acknowledgments}  --
We are grateful to T.~Hahn, J.-L.~Kneur, and A.~Pukhov for patient
correspondence on their computer codes.
Furthermore, we are grateful to T.~Plehn, M.~Pospelov, and Y.Y.Y.~Wong
for valuable discussions.
This research was supported in part by the DFG cluster of excellence
``Origin and Structure of the Universe.''

\bigskip

{\bf Note added} -- Ref.~\cite{Ratz:2008qh}, in which the potential
suppression in the stau yield $\Ystau$ due to an enhanced annihilation
into $\hhiggs\hhiggs$ final states is also studied, appeared as this
work was being finalized.
This paper provides analytic approximations for the stau annihilation
cross section into $\hhiggs\hhiggs$ and for the associated yield. In
addition, results of numerical studies within the CMSSM, the NUHM, and
a scenario with non-universal gaugino masses are presented that
exhibit parameter regions with extremely small $\Ystau$ values.
In our work also enhanced stau annihilation into $\hhiggs\Hhiggs$ and
into $\Hhiggs\Hhiggs$ and stau annihilation at the $\Hhiggs$
resonance, which were not considered in~\cite{Ratz:2008qh}, are discussed.
In addition, our work provides a systematic investigation based on a
complete set of stau annihilation channels, an outline of the way in
which the mechanisms leading to the suppression of $\Ystau$ can be
probed at collider experiments, and a thorough presentation of the
potential implications for gravitino dark matter scenarios.

\bibliography{biblio}

\end{document}